\documentclass[12pt]{article}
\usepackage{latexsym,amssymb}
\newcommand{\be}{\begin{equation}}
\newcommand{\ee}{\end{equation}}
\newcommand{\beqs}{\begin{eqnarray}}
\newcommand{\eeqs}{\end{eqnarray}}
\def\vv{{\cal V}}
\def\({\left(}
\def\){\right)}

\newcommand{\Exc}[1]{{${\rm E}_{{#1}({#1})}$}}

\def\mxth{\mathsurround=0pt }
\def\xversim#1#2{\lower2.pt\vbox{\baselineskip0pt \lineskip-.5pt
x  \ialign{$\mxth#1\hfil##\hfil$\crcr#2\crcr\sim\crcr}}}

\def\slash{\llap /}

\renewcommand{\a}{\alpha}
\renewcommand{\b}{\beta}

\renewcommand{\d}{\delta}
\newcommand{\pa}{\partial}
\newcommand{\g}{\gamma}

\newcommand{\e}{\epsilon}

\newcommand{\m}{\mu}
\newcommand{\n}{\nu}

\newcommand{\nn}{\nonumber}

\def\be{\begin{equation}}
\def\ee{\end{equation}}
\def\bea{\begin{eqnarray}}
\def\eea{\end{eqnarray}}

\newcommand{\ft}[2]{{\textstyle\frac{#1}{#2}}}

\newcommand{\eqn}[1]{(\ref{#1})}
\def\bfone{\relax{\rm 1\kern-.35em 1}}

\begin{document}
\begin{titlepage}

\begin{center}
ITP-UU-04/31 \qquad\quad SPIN-04/18 \qquad\quad DESY 04-245
\end{center}
\vskip 12mm
\begin{center}
{\Large {\bf THE MAXIMAL D$=$5 SUPERGRAVITIES}}
\end{center}
\vskip 6mm

\begin{center}
{{\bf Bernard de Wit}\\
 Institute for Theoretical Physics \,\&\, Spinoza Institute,\\  
Utrecht University, Postbus 80.195, NL-3508 TD Utrecht, 
The Netherlands\\
{\tt b.dewit@phys.uu.nl}}
\vskip 2mm
{{\bf Henning Samtleben}\\
II. Institut f\"ur Theoretische Physik der Universit\"at Hamburg,\\ 
 Luruper Chaussee 149, D-22761 Hamburg, Germany\\
{\tt henning.samtleben@desy.de}}
\vskip 2mm
{{\bf Mario Trigiante}\\
Dept. of Physics, Politecnico di Torino,\\
Corso Duca degli Abruzzi 24, I-10129 Torino, Italy \\
{\tt mario.trigiante@to.infn.it}}
\vskip 4mm

\end{center}

\vskip .2in

\begin{center} {\bf Abstract } \end{center}
\begin{quotation}\noindent
The general Lagrangian for maximal supergravity in five spacetime dimensions
is presented with vector potentials in the $\overline{\bf 27}$ and tensor
fields in the ${\bf 27}$ representation of \Exc6. This novel tensor-vector
system is subject to an intricate set of gauge transformations, describing
$3(27-t)$ massless helicity degrees of freedom for the vector fields and $3t$
massive spin degrees of freedom for the tensor fields, where the (even)
value of $t$ depends on the gauging. The kinetic term
of the tensor fields is accompanied by a unique Chern-Simons coupling which
involves both vector and tensor fields. The Lagrangians are completely encoded
in terms of the embedding tensor which defines the \Exc6 subgroup 
that is gauged by the vectors. 
The embedding tensor is subject to two constraints which ensure the
consistency of the combined vector-tensor gauge transformations and the
supersymmetry of the full Lagrangian. This new formulation encompasses all
possible gaugings. 
\end{quotation}
\end{titlepage}
\eject
\section{Introduction}
Maximal supergravities without gauging can be formulated on the basis of
different field representations via tensor dualities which convert
antisymmetric tensor fields of rank $p$ into tensor fields of 
rank $D-p-2$. The choice of the field representation has implications for the
symmetry group of the Lagrangian, but not of the field equations. Therefore
this issue has a bearing on the introduction of possible gauge  
interactions associated with a nontrivial gauge group, as this group should be
embedded into the symmetry group of the Lagrangian. In fact the situation is
even more subtle, as gauge interactions with charged `matter' fields may be
incompatible with other independent gauge invariances that these fields may be
subject to. 

In five spacetime dimensions vector and tensor gauge fields are dual to
one another in the absence of charges. The ungauged maximal supergravity
Lagrangian is described in terms of  vector fields transforming according to
the $\overline{\bf 27}$ representation  
of \Exc6  \cite{cremmer80}, where the presence of an abelian Chern-Simons term
forms an obstacle to dualizing all vectors into tensor gauge
fields. Obviously, a partial dualization destroys the manifest \Exc6
invariance of the Lagrangian. When 
switching on a gauging, the corresponding gauge fields transform according to 
the adjoint representation of the gauge group. The dimension of the gauge
group is usually less than 27, so that there are vector fields that do not
belong to this adjoint representation. When these gauge fields carry charges
that cannot be incorporated into a central extension of the gauge algebra,
then the gauging can only exist provided these fields can be 
converted to charged tensor fields. The corresponding tensor field Lagrangians
have a kinetic term linear in 
spacetime derivatives and proportional to a five-dimensional
Levi-Civita tensor, which allows a minimal coupling to gauge fields and a
possible mass term. Indeed, this option was exploited in
\cite{GunaRomansWarner}, where the gauging of maximal five-dimensional
supergravity with the 15-dimensional gauge groups ${\rm SO}(q,6-q)$ was
constructed. In that case there are 12 massive charged tensor fields,
transforming in the $({\bf 6},{\bf 2})$ representation of 
${\rm SO}(q,6-q)\times {\rm SL}(2,\mathbb{R})$. Prior to that work a similar
situation had already been noted in seven spacetime dimensions \cite{PPvN}
(see also \cite{TownPilcNieuw}). 

The above seems to imply that, in five spacetime dimensions, a gauging cannot
just be effected by switching on the gauge charges, as the field
representation must first be suitably adapted. This feature has hampered a
general analysis of all possible gaugings. One of the central results of this 
paper is a new formulation of five-dimensional maximal supergravity that is
sufficiently flexible to incorporate all necessary field representations from
the start, thus enabling a general analysis of all possible gaugings. So far,
the known gaugings \cite{GunaRomansWarner,AndCordFreGual,dWST1}
comprise  the ${\rm SO}(q,6-q)$ gaugings, contractions thereof, and gaugings 
induced by reduction from higher-dimensional
supergravities (although few of those have been discussed
in detail). In the context of the AdS/CFT correspondence the ${\rm SO}(6)$
gauging received most attention.   

Beyond these typical five-dimensional issues, one must address the
modification of the Lagrangian with masslike terms and a scalar
potential. These new couplings are encoded in 
the so-called $T$-tensor \cite{deWitNic}, and recently it was demonstrated
how viable gaugings can be investigated by means of a group-theoretical
analysis of this $T$-tensor \cite{dWST1,dWST2,dWST3}. The purpose
of this paper is to demonstrate how this analysis leads to a completely
general treatment of all maximal gauged supergravities, using the new
formulation of the Lagrangian mentioned above. This formulation is based on
vector fields and tensor fields transforming in the $\overline{\bf 27}$ and
${\bf 27}$ representations of \Exc6, respectively, and this combined system of
vector and tensor fields is subject to both vector and 
tensor gauge invariances encoded in the same embedding tensor that
determines the gauge group and the $T$-tensor. This intricate gauge invariance
guarantees that the combined system describes always 81 degrees of freedom, as
required by supersymmetry. The embedding tensor is treated
as a spurionic object which transforms under \Exc6, which makes it amenable to
a group-theoretical analysis. The Lagrangian remains formally \Exc6
invariant until the embedding tensor is frozen to a constant. The embedding
tensor is subject to two constraints: it must belong to the ${\bf 351}$
representation of \Exc6, and the $\overline{\bf 27} + {\bf 1728}$
representation contained in its square should cancel. For any embedding tensor
that satisfies these constraints there exists a consistent supersymmetric and
gauge invariant Lagrangian. 

The relation between the embedding tensor and the $T$-tensor involves the
representative of the ${\rm E}_{6(6)}/{\rm USp}(8)$ coset space that is
parametrized by the scalar fields. Here we use the standard
treatment of gauged nonlinear sigma models in which the group ${\rm USp}(8)$
is realized as a local invariance which acts on 
the spinor fields and the scalars; the corresponding connections
are composite fields. A gauging is based on a group
${\rm G}_g\subset {\rm E}_{6(6)}$ whose connections are (some of the)
elementary vector gauge fields of the supergravity
theory. The coupling constant associated with the gauge group ${\rm
G}_g$ will be denoted by $g$. One can impose a gauge
condition with respect to the local ${\rm USp}(8)$ invariance which amounts to 
fixing a coset representative for the coset space. In that case the
\Exc6-symmetries will act nonlinearly on the fields and these
nonlinearities make many calculations intractable or, at
best, very cumbersome. Because it is much more convenient to work with
symmetries  that are realized linearly, the best strategy is therefore
to postpone the gauge fixing till the end.

The new Lagrangian based on the combined vector-tensor gauge invariance and its
supersymmetry transformations are universal in the sense that they will take
the same form irrespective of the gauging. All the details of the gauging are
encoded into the embedding tensor and the quantities related to it. Once the
group-theoretical constraints on the embedding tensor are satisfied, it is
guaranteed that the gauging is consistent with supersymmetry. Hence our
results encompass all possible maximal supergravity theories in five
dimensions.  

This paper is organized as follows. In section~2 we discuss the embedding
tensor and the constraints it must satisfy. This analysis motivates the new
formulation of the Lagrangian with the combined vector-tensor system, which is
presented in section~3. In 
section~4 we define the $T$-tensor and derive the consequences of the
group-theoretical constraints. This requires a detailed discussion of the
characteristic features of the ${\rm E}_{6(6)}/{\rm USp}(8)$ coset
space.  In section~5 we discuss the Lagrangian and the supersymmetry
transformation rules, up to higher-order fermion terms. Finally, in section~6
we analyze a number of examples pertaining to known and new gaugings, 
and in section~7 we present our concluding remarks. 

\section{The embedding tensor}
\setcounter{equation}{0}
\label{embed}
The (abelian) vector fields $A_\m{}^M$ transform in a representation 
$\overline{\bf 27}$ of \Exc6 with generators denoted by 
$(t_\a)_M{}^N$, so that $\d A_\m{}^M = -\Lambda^\a (t_\a)_N{}^M\,A_\m{}^N$.
The gauge group is a subgroup of \Exc6 so that the generators $X_M$ 
are decomposable in terms of the 78 independent \Exc6-generators $t_\a$, {\it
  i.e.},   
\be
\label{X-theta-t}
X_M = \Theta_M{}^\a\,t_\a\;,
\ee
where $\a=1,2,\ldots,78$ and $M=1,2,\ldots,27$. 
The gauging is thus encoded in a real {\it embedding tensor}
$\Theta_{M}{}^{\alpha}$ assigned to the ${\bf 27}\times{\bf 78}$
representation of \Exc6. The embedding tensor acts as
a projector whose rank $s$ equals the dimension of the gauge group (not
counting abelian gauge fields corresponding to a central extension of the
gauge algebra, as we will discuss in due course). 
The strategy of this paper is to treat the embedding tensor as a spurionic
object that transforms under \Exc6, so that the Lagrangian and 
transformation rules remain formally \Exc6 covariant. The embedding
tensor can then be characterized group-theoretically. When freezing
$\Theta_M{}^\a$ to a constant, the \Exc6-invariance is broken. 
An admissible embedding tensor is subject to a linear and a quadratic
constraint, which 
ensure that one is dealing with a proper subgroup of \Exc6 and that the
corresponding supergravity action remains supersymmetric. These constraints
are derived in the first subsection. The second subsection describes a number
of implications of these constraints, while a third subsection presents some
of the results in a convenient basis. 
\subsection{The constraints on the embedding tensor}
The fact that the $X_M$ generate a group and thus define a Lie algebra,
\be
\label{gauge-algebra}
{[X_M,X_N]} = f_{MN}{}^P\,X_P,
\ee
with $f_{MN}{}^P$ the as yet unknown structure constants of the gauge group,
implies that the embedding tensor must satisfy the closure condition,
\be
\label{gauge-gen}
\Theta_M{}^\a\,\Theta_N{}^\b \,f_{\a\b}{}^{\g}= f_{MN}{}^P\,
\Theta_P{}^\g\,.
\ee
Here the $f_{\a\b}{}^\g$ denote the structure constants of \Exc6,
according to $[t_\a,t_\b]= f_{\a\b}{}^\g\,t_\g$. The
closure condition implies that the structure constants $f_{MN}{}^P$ satisfy the
Jacobi identities in the subspace projected by the embedding tensor,
\be
\label{jacobi}
f_{[MN}{}^Q\,f_{P]Q}{}^R\,\Theta_R{}^\a =0\,.
\ee

Once the gauge group is specified, one introduces covariant derivatives 
given by  
\be
D_\m = \partial_\m - g\,A_\m{}^M\,X_M\,, 
\ee
where $g$ denotes the gauge coupling constant. They lead to the covariant
field strengths, 
\be
\Theta_M{}^\a \,{\cal F}_{\m\n}{}^M = \Theta_M{}^\a (\pa_\m A_\n{}^M  - \pa_\n
A_\m{}^M   -g\,f_{NP}{}^M \, A_\m{}^N\, A_\n{}^P)\,.  
\ee
The gauge field transformations are given by
\be
\Theta_M{}^\a \,\d A_\m{}^M = \Theta_M{}^\a\, (\pa_\mu \Lambda^M - g\,
f_{NP}{}^M \,A_\m{}^N\,\Lambda^P)\,.
\ee
Because of the contraction with the embedding tensor, the above results only
apply to an $s$-dimensional subset of the gauge fields; the remaining ones do
not appear in the covariant derivatives and are not directly involved in the
gauging. 
However, the $s$ gauge fields that do appear in the covariant derivatives, 
are only determined up to additive terms linear in the $27-s$ gauge fields
that vanish upon contraction with $\Theta_M{}^\a$. 

While the gauge fields involved in the gauging should transform in the adjoint
representation of the gauge group, the gauge field charges should also
coincide with $X_M$ in the $\overline{\bf 27}$ representation. Therefore
$(X_M)_N{}^P$  must decompose into the adjoint representation of the gauge 
group plus possible extra terms which vanish upon contraction with the
embedding tensor, 
\be
\label{adjoint}
(X_M)_N{}^P \,\Theta_P{}^\a \equiv \Theta_M{}^\b \,t_{\b N}{}^P\,
 \Theta_P{}^\a  = - f_{MN}{}^P \,\Theta_P{}^\a \,.
\ee
Note that \eqn{adjoint} is the analogue of \eqn{gauge-gen} in the
$\overline{\bf 27}$ representation.  
The combined conditions \eqn{gauge-gen} and \eqn{adjoint} imply that $\Theta$
is invariant under the gauge group and yield the \Exc6-covariant condition
\begin{equation}
  \label{eq:closure-constraint}
  C_{MN}{}^\a \equiv f_{\b\g}{}^\a \, \Theta_M{}^\b\,\Theta_N{}^\g +
t_{\b N}{}^P\,
\Theta_M{}^\b \, \Theta_P{}^\a =0 \;.
\end{equation}

Obviously $C_{MN}{}^\alpha$ can be assigned to irreducible \Exc6
representations contained in the 
${\bf 27} \times {\bf 27}\times{\bf 78}$ representation. The
condition \eqn{eq:closure-constraint} encompasses all previous results: it
implies that  
\be
\label{eq:X-closure}
{[X_M,X_N]} = -X_{MN}{}^P\,X_P,
\ee
so that \eqn{eq:closure-constraint} implies a closed gauge
algebra, whose structure constants, related to $X_{MN}{}^P$ in accord with
\eqn{adjoint}, have the required antisymmetry. Hence
\eqn{eq:closure-constraint} 
is indeed sufficient for defining a proper subgroup embedding.\footnote{
 Note that for an abelian gauge group we have
 $X_{MN}{}^P\Theta_P{}^\a=0$. Using \eqn{eq:repres-constraint} this leads to
 ${\rm tr}(X_M\,X_N) =0$.}  

The embedding tensor satisfies a second constraint, which is required by 
supersymmetry. This constraint is linear and restricts the
embedding tensor to the ${\bf 351}$ representation \cite{dWST1}. From  
\be
{\bf 27} \times {\bf 78}=  {\bf 27}+{\bf 351}+\overline{{\bf 1728}} \,,
\ee
one shows that this condition on the representation implies the equations, 
\begin{equation}
    \label{eq:repres-constraint}
    t_{\alpha M}{}^N\,\Theta_N{}^\alpha = 0\,,\qquad
   (t_{\beta} t^{\alpha})_ M{}^N\,\Theta_N{}^\beta = -\ft23\,
    \Theta_M{}^\alpha \,,
\end{equation}
where the index $\a$ is raised by the inverse of the \Exc6-invariant
metric $\eta_{\alpha\beta}= {\rm tr}(t_\a t_\b)$. 

As a result of this constraint, the representation content of
$C_{MN}{}^\alpha$ can be further restricted. From 
\eqn{eq:repres-constraint} one can derive the following equations,
\begin{equation}
  \label{eq:C-constraints}
t_{\alpha N}{}^P\, C_{MP}{}^\alpha =0 \,, \quad
(t_\beta \,t^{\alpha})_ N{}^P\, C_{MP}{}^\beta =
- \ft23\,C_{MN}{}^\alpha \,,\quad
t_{\alpha M}{}^P\, C_{PN}{}^\alpha= t_{\alpha N}{}^P\,
C_{PM}{}^\alpha\,.
\end{equation}
They imply that $C_{MN}{}^\alpha$ belongs to representations
contained in ${\bf 27}\times{\bf 351}$. On the other hand, the product of two
$\Theta$-tensors belongs to the symmetric product of two ${\bf 351}$
representations. Comparing the decomposition of these two
products\footnote{
  We used the LiE package \cite{LeCoLi92} for computing the decompositions of
  tensor products and the branching of representations.}, 
\begin{eqnarray}
\label{square}
({\bf 351}\times {\bf 351})_{\rm s} &=&
\overline{{\bf 27}}+{\bf 1728}+
{\bf 351}{}^\prime+{\bf 7722}+{\bf 17550}+{\bf 34398} \,, \nonumber\\ 
{\bf 27}\times {\bf 351}&=& 
\overline{{\bf 27}}+{\bf 1728}+ \overline{{\bf 351}}+{\bf 7371}\,, 
\end{eqnarray}
one deduces that $C_{MN}{}^\a$ belongs to the $\overline{\bf 27}+{\bf 1728}$ 
representation.

Summarizing, a consistent gauging is defined by an embedding tensor
$\Theta_M{}^\alpha$ satisfying the linear constraint
(\ref{eq:repres-constraint}) together with the
quadratic constraint (\ref{eq:closure-constraint}) with the \Exc6
representation content,  
\begin{eqnarray}
  \label{eq:sum}
  \Big({\mathbb{P}}_{\bf 27}  +{\mathbb{P}}_{\overline{\bf 1728}}\Big)
  \,\Theta  &=&   0\,,   \nn\\ 
\Big( {\mathbb{P}}_{\overline{\bf 27}} + {\mathbb{P}}_{\bf 1728}
  \Big)\,\Theta\,   \Theta &=&  0\,.  
\end{eqnarray}
\subsection{Some implications of the embedding tensor constraints}
Because $(t_\a)_M{}^N$ is an \Exc6-invariant tensor, it follows that
$X_{MN}{}^P$ transforms in the ${\bf 351}$ representation of \Exc6, just as
the embedding tensor. Furthermore 
the product of three ${\bf 27}$ representations contains a singlet
representation, associated with a symmetric \Exc6-invariant tensor
$d_{MNP}$. The same is true for the conjugate representation, so that there
exists also a symmetric invariant tensor $d^{MNP}$. Hence it follows that
$X_{MN}{}^P$ satisfies the following properties, 
\be
\label{X-prop}
X_{MN}{}^N=X_{NM}{}^N=0 \,, \qquad X_{M(N}{}^R d_{PQ)R}=0 = X_{MN}{}^{(P}
d^{QR)N} \,. 
\ee
By writing $X_{MN}{}^P=X_{(MN)}{}^P+X_{[MN]}{}^P$, it seems that one can
decompose the tensor $X_{MN}{}^P$ into two representations, while, on the other
hand, we know that $X_{MN}{}^P$ must belong to a single irreducible
representation. Therefore both the symmetric and the antisymmetric
components should be proportional to the same tensor transforming in the 
${\bf 351}$ representation. This is confirmed by the fact that 
the $\overline{\bf 27}$ representation yields a ${\bf 351}$ representation
when multiplied with either the symmetric or the antisymmetric product of two
${\bf 27}$ representations,
\bea
\label{mixed-cubic-27}
\overline{\bf 27} \times({\bf 27}\times {\bf 27})_{\rm s} &=& \overline{\bf
  27} \times(\overline{{\bf 27}}+{\bf 351}^\prime) = {\bf 351}+{\bf 27}+ {\bf
  27}+ \overline{{\bf 351}}^\prime +\overline{{\bf 1728}}+\overline{{\bf
    7722}}\,, \nonumber\\
\overline{\bf 27} \times({\bf 27}\times {\bf 27})_{\rm a} &=&
 \overline{{\bf 27}}\times \overline{{\bf 351}} = {{\bf 351}}+{{\bf 27}}
 +\overline{\bf 1728}+\overline{\bf 7371}\,.
\eea
Therefore we can construct two contractions of $X_{MN}{}^P$
with invariant tensors yielding a tensor $Z^{MN}$ that must be antisymmetric
so that it transforms in the ${\bf 351}$ representation. In both cases we
should find the same tensor, {\it i.e.},  
\begin{eqnarray}
\label{eq:a-sym-X}
X_{PQ}{}^M\, d^{NPQ}&=& Z^{MN}\,, \nonumber\\
2\,X_{PQ}{}^T\, d^{PRM}d^{QSN}d_{RST}&=& Z^{MN}\,.
\end{eqnarray}
We observe that the antisymmetry in $[MN]$ of the first equation of
\eqn{eq:a-sym-X} follows also from  \eqn{X-prop}.
Possible additional proportionality factors in \eqn{eq:a-sym-X} can be absorbed
into the invariant tensors $d_{MNP}$ and $d^{MNP}$.  Using \eqn{X-prop} can show that the  
factors in \eqn{eq:a-sym-X} are consistent provided that we choose the
relative normalizations of the two tensors such that
\begin{equation}
  \label{eq:d-normalization}
d_{MPQ}\,d^{NPQ}= \d_M{}^N\,.  
\end{equation}
Because the symmetric product of four ${\bf 27}$ representations contains
precisely one ${\bf 27}$ representation, we deduce another identity, 
\begin{equation}
  \label{eq:3-d-identity}
d_{S(MN}\, d_{PQ)T}\, d^{STR} = \ft2{15}\,\d^R{}_{\!\!(M}\,d_{NPQ)}\,.  
\end{equation}
Contraction with additional invariant tensors yields a number of other useful
identities.\footnote{
  The following identities proved convenient:
\bea
d_{MRS}\,d^{SPT} \, d_{TNU} \,d^{URQ} &=& \ft1{10}\,\d_{(MN)}{}^{\!\!(PQ)}  - 
\ft25\, d_{MNR}\, d^{PQR}\,, \nonumber \\
d_{MPS}\,d^{SQT} \, d_{TRU} \,d^{UPV}\,d_{VQW}\,d^{WRN}  &=& - \ft3{10} 
\,\d_{M}{}^{N}\,.  \nn
\eea  }  
With these results one can derive the inverse relations of \eqn{eq:a-sym-X}, 
\begin{equation}
\label{X-dZ}
X_{(MN)}{}^P = d_{MNQ}\,Z^{PQ}  \,,\qquad
X_{[MN]}{}^P = 10\,d_{MQS}\, d_{NRT}\, d^{PQR}\, Z^{ST} \,.
\end{equation}

A number of important identities quadratic in the embedding tensor can also be
derived. The first one concerns the expression
$Z^{MN}\Theta_N{}^\a$. This tensor transforms in the representation,
$\overline{\bf 27} \times {\bf 78}=  \overline{\bf 27}+\overline{\bf 351}+
{{\bf 1728}}$, 
and  can be compared to the square of the embedding tensor, which yields the
representations listed in the first equation \eqn{square}. The only
representation they have in common, however, is the  
$\overline{\bf 27}+{{\bf 1728}}$, which vanishes because of
the constraint \eqn{eq:closure-constraint}. Therefore, we conclude, 
\begin{equation}
\label{eq:Z-X-orthogonal}
Z^{MN}\,\Theta_N{}^\a = 0\,,\qquad Z^{MN}\,X_{N}= 0\,,
\end{equation}
where, in the second equation, $X_M$ is taken in an arbitrary representation. 
Along the same lines, we can consider the contraction
$X_{MN}{}^{[P}\,Z^{Q]N}$. From the second
branching of \eqn{mixed-cubic-27}, we readily deduce that this tensor
should belong to the 
$\overline{{\bf 351}}+\overline{{\bf 27}} +{\bf 1728}+{\bf 7371}$
representation. Comparing these representations to those generated by the
square of the embedding tensor ({\it c.f.} the first equation \eqn{square}) we
note that the only representations they have in common are again the ones
which are set to zero by the constraint \eqn{eq:closure-constraint}. Hence the
tensor $Z^{MN}$ is {\it invariant} under the gauge group,
\begin{equation}
\label{eq:Z-invariance}
X_{MN}{}^{[P}\,Z^{Q]N}= 0\,. 
\end{equation}

The reasoning that led to (\ref{eq:Z-X-orthogonal}) and
(\ref{eq:Z-invariance}) can be applied to show that we are in fact dealing
with equivalent forms of the quadratic closure constraint
(\ref{eq:closure-constraint}), at least for
embedding tensors that are restricted to the ${\bf 351}$ representation. We
list three equivalent forms of the quadratic constraint, 
\begin{eqnarray}
  \label{eq:equiv-quadr}
  X_{MP}{}^R\,X_{NR}{}^Q -X_{NP}{}^R\,X_{MR}{}^Q + X_{MN}{}^R\,X_{RP}{}^Q &=&
  0 \,,\nonumber\\
  Z^{MN}\,X_{N}&=& 0\,, \nonumber\\
  X_{MN}{}^{[P}\,Z^{Q]N}&=& 0\,. 
\end{eqnarray}

In the next section we will use the tensor $X_{[MN]}{}^P$ as an extension of
the gauge group structure constants $f_{MN}{}^P$, 
which satisfies the Jacobi identity up to terms proportional to $Z$. From
\eqn{eq:X-closure} and some of the previous identities one derives,
\begin{eqnarray}
\label{Jacobi-X}
 \lefteqn
 {X_{[MN]}{}^P\, X_{[QP]}{}^R + X_{[QM]}{}^P\, X_{[NP]}{}^R  + X_{[NQ]}{}^P \,
 X_{[MP]}{}^R}   \nonumber \\   
 && \qquad\qquad{}= \Big\{d_{SPQ}\, X_{[MN]}{}^P + 2\, d_{SP[M}\, d_{N]QO}\, Z^{OP}\Big\}
 Z^{SR} \nonumber \\   
 && \qquad\qquad{}=d_{SP[Q}\, X_{MN]}{}^P\, Z^{SR}\,.  
\end{eqnarray}
As the right-hand side vanishes upon contraction with the embedding tensor
$\Theta_R{}^\a$, we see that the $X_{[MN]}{}^P$ satisfy the Jacobi identity
in the subspace projected by the embedding tensor, just as the gauge group
structure constants ({\it c.f.} \eqn{jacobi}). 
\subsection{A special \Exc6 basis}
\label{specialbasis}
Let us now consider a basis for the vector fields such that all the nonzero
components of the 78 vectors $\Theta^\a$ cover an $s$-dimensional subspace
parametrized by the gauge fields $A_\m{}^M$ with $M=1,\ldots,s$ and $s\leq
27$. In this basis $X_{MN}{}^P$ can be written in triangular form,
\be
\label{blockmatrix}
X_M= \pmatrix{ - f_M & a_M\cr \noalign{\vskip 1mm}0& b_M\cr}\,,
\ee
where the $s\times s$ upper-left diagonal block coincides with the gauge group
structure constants and the contribution of the submatrices $a_M$ and $b_M$
vanish in the product $(X_M)_N{}^P\Theta_P{}^\a$. The lower-left $s\times
(27-s)$ block vanishes as a result of \eqn{adjoint}. 
It is easy to see that $a_M$ and $b_M$ cannot both be zero. If that were the
case, 
we would have $f_{MN}{}^P = - X_{MN}{}^P$, which is antisymmetric in $M$ and
$N$. Hence,  
\be
\label{eq:antisymmetry}
\Theta_N{}^\alpha\,t_{\alpha\,M}{}^P=
- \Theta_M{}^\alpha\,t_{\alpha\,N}{}^P\,.
\ee
Contracting this result by $(t^\beta)_P{}^M$ leads to $t_\alpha\,t^\beta \,
\Theta^\alpha = - \Theta^\beta$ which is in contradiction with the
representation constraint \eqn{eq:repres-constraint}. 

Let us now refine this choice of basis and consider some of the results of the
previous subsection. In this special \Exc6 basis the
components of a vector $V_M$ belonging to the ${\bf 27}$ representation 
are decomposed as $V_M=(V_A,\,V_a,\,V_u)$, where $A=1,\ldots,s$, $a=
s+1,\ldots,27-t$ and $u= 28-t, \ldots, 27$, where the only nonvanishing
components of  
$\Theta_M{}^\a$ are $\Theta_A{}^\a$ and therefore the $X_{MN}{}^P$ are nonzero
only if $M=A$. Correspondingly we decompose the vector $V^M$ transforming in
the $\overline{\bf 27}$ representation as $V^M=(V^A,\,V^a,\,V^u)$. Obviously
the tensor $Z^{MN}$ vanishes whenever $M$ or $N$ are equal to $A, B,\ldots$
in view of \eqn{eq:Z-X-orthogonal} and the distinction between indices $a,
b,\ldots$ and $u,v,\ldots$ is due to the fact that we assume that only
$Z^{uv}$ is nonvanishing. Hence, $V^a$ and $V^u$ span the subspace orthogonal
to $\Theta_M{}^\a$ and $V_A$ and $V_a$ span the subspace orthogonal to
$Z^{uv}$. Consequently, the number $t$ of indices $u,v$ must be {\it even}.
{}From \eqn{X-dZ},~\eqn{eq:Z-X-orthogonal} and \eqn{eq:Z-invariance}, it follows
that $X_{AN}{}^P$ has a block decomposition, which goes beyond 
\eqn{blockmatrix} (row and column indices are denoted by $B,b,v$ and $C,c,w$,
respectively), 
\be
\label{X-decomposition} 
X_{AN}{}^P = \pmatrix{ -f_{AB}{}^C & h_{AB}{}^c & C_{AB}{}^w \cr
\noalign{\vskip 1mm}
                         0         & 0          & C_{Ab}{}^w \cr 
\noalign{\vskip 1mm}
                         0         & 0          & D_{Av}{}^w \cr}\;,
\ee
where we note the following relations,
\be
h_{(AB)}{}^c = f_{(AB)}{}^C = f_{AB}{}^B= D_{Au}{}^u= Z^{w[u}\,D_{Aw}{}^{v]}
=0\,, 
\ee
which imply that the gauge group is unimodular. Observe that the basis choice
for $V^M$ and $V_M$ is not unambiguous. The $V^A$ can still be modified by
linear combinations of $V^a$ and $V^u$, and likewise, the $V_u$ can be
modified by terms linear in $V_A$ and $V_a$. These redefinitions do not alter 
the general form of \eqn{X-decomposition} but they affect the expressions for
the nondiagonal blocks. 

{}From other relations derived above (in particular from the first equation
\eqn{X-dZ}), we can establish a variety of results, 
\bea
C_{(AB)}{}^u &=& d_{ABv}\,Z^{uv} \,, \nonumber \\
C_{Aa}{}^u &=& 2\,d_{Aav}\,Z^{uv} \,, \nonumber \\
D_{Av}{}^u &=& 2\,d_{Avw}\,Z^{uw} \,.
\eea
Furthermore one derives that $d_{uvw} =d_{uva}= d_{uab}=0$. Other
identities follow from the invariance of $d_{MNP}$, such as
\bea
 f_{AB}{}^C \, d_{abC} + 4\, Z^{wx}\, d_{Aw(a}\,d_{b)xB}&=& 0\,, \nonumber\\
 f_{AB}{}^C \, d_{uvC} + 4\, Z^{wx}\, d_{Aw(u}\,d_{v)xB}&=& 0\,, \nonumber\\
 f_{AB}{}^C \, d_{auC} + 2\, Z^{wx}\, d_{Awa}\,d_{uxB}+ 2\, Z^{wx}\,
 d_{Awu}\,d_{axB}   &=& 0\,.
\eea
Likewise, the invariance of the $d^{MNP}$ tensor leads to a large variety of
equations, of which we present the following two,
\bea
f_{AE}{}^{(B}\, d^{CD)E} &=& 0\,, \nonumber \\
h_{AD}{}^a\, d^{BCD} -2\, f_{AD}{}^{(B}\, d^{C)Da} &=& 0\,.
\eea
The closure relations \eqn{gauge-algebra} imply three additional identities, 
\bea
\label{group-ids}
f_{[AB}{}^D\,f_{C]D}{}^E &=&0\,,\nonumber \\
f_{[AB}{}^D\,h_{C]D}{}^a &=&0\,,\nonumber \\
2\,f_{C[A}{}^D\,C_{B]D}{}^u  -f_{AB}{}^D\,C_{DC}{}^u 
 - 2\, h_{C[A}{}^a\,C_{B]a}{}^u &=&0\,.
\eea
{}From the first two equations it follows that the upper-left submatrix of
\eqn{X-decomposition} parametri\-zed in terms of $f_{AB}{}^C$ and $h_{AB}{}^a$
closes under commutation and defines consistent gauge transformation rules for
the gauge fields $A_\m{}^A$ and $A_\m{}^a$, 
\bea
\d A_\m{}^A &=& \pa_\mu \Lambda^A - g\,
f_{BC}{}^A \,A_\m{}^B\,\Lambda^C \,, \nonumber\\
\d A_\m{}^a &=& \pa_\mu \Lambda^a + g\,
h_{BC}{}^a \,A_\m{}^B\,\Lambda^C \,.
\eea
The corresponding field strengths read, 
\bea
\label{cov-field-strengths}
{\cal F}_{\m\n}{}^A &=& \pa_\m A_\n{}^A  - \pa_\n
A_\m{}^A   -g\,f_{BC}{}^A \, A_\m{}^B A_\n{}^C \,,  \nonumber\\
{\cal F}_{\m\n}{}^a &=& \pa_\m A_\n{}^a  - \pa_\n
A_\m{}^a   + g\,h_{BC}{}^a \, A_\m{}^B A_\n{}^C \,.  
\eea
The only gauge fields that appear in the covariant derivatives
are the fields $A_\m{}^A$, so no other gauge fields couple to charges that act
on the matter fields. However, to write consistent transformation rules for
the gauge fields, one must incorporate the abelian gauge fields $A_\m{}^a$
into the gauge algebra. These gauge fields couple to charges that are central
in the gauge algebra so that the gauge algebra is a central
extension of the algebra \eqn{gauge-algebra}. Introducing formal generators
$\tilde X_A$ and $\tilde X_a$, it reads,
\be
{[}\tilde X_A,\tilde X_B]= f_{AB}{}^C\,\tilde X_C - h_{AB}{}^a \tilde X_a\,. 
\ee
On the matter fields the charges $\tilde X_a$ vanish and the gauge algebra
coincides with \eqn{gauge-algebra}. 

The remaining gauge fields $A_\m{}^u$ carry charges related to the last
column in \eqn{X-decomposition}. Since these charges cannot be incorporated in
the gauge transformations on the vector fields and lead to inconsistent
couplings, these gauge fields must be dualized to charged tensor fields. This
is a well-known feature in gauged supergravities in odd dimensions 
\cite{GunaRomansWarner,PPvN,TownPilcNieuw}, as we already discussed in the
introduction. Therefore the physically relevant gauge group is the one
associated with the gauge fields $A_\m{}^A$ and $A_\m{}^a$.

However, to dualize vectors to tensors affects the manifest \Exc6 covariance 
of our results, and requires a case-by-case dualization for every separate
gauging. Therefore, rather than to perform this dualization, we will keep all
27 gauge fields and introduce 27 tensor fields as well. At the same time we
introduce an extended tensor-vector gauge invariance in order to balance the
degrees of freedom. Upon a suitable gauge condition one can remove
the gauge fields $A_\m{}^u$ whose degrees of freedom are then carried by the
tensor fields. This implies that we must introduce a new gauge transformation
for the vector fields of the form $\delta A_\m{}^M\propto 
Z^{MN}\,\Xi_{\m\,N}$ 
where $\Xi_{\m\,N}$ is the gauge parameter, that enables to remove the gauge
fields so that they are effectively replaced by the tensor fields. When the
gauge-fixing is postponed until the end the Lagrangian for this vector-tensor
system takes a unique and \Exc6-covariant form. The gauging is encoded in
terms of the embedding tensor, or equivalently, in terms of the tensor
$Z^{MN}$. This formulation is discussed in the next section. 

\section{Tensor and vector gauge fields}
\setcounter{equation}{0}
\label{vector-tensor}
\newcommand{\X}[2]{X_{#1}{}^{#2}}
As explained in the introduction we will set up a formulation based on vector
and tensor fields, $A_\m{}^M$ and $B_{\m\n\,M}$, which transform in the
$\overline {\bf 27}$ and ${\bf 27}$ representation of \Exc6,
respectively. The combined vector and tensor gauge transformations will ensure
that the number of physical degrees of freedom will remain independent of the
embedding tensor. The latter will only determine how the degrees of freedom
are shared between the vector and tensor fields. Two ingredients play a
crucial role in order to accomplish this. First of all the vector fields
transform under tensor gauge transformations proportional to $Z^{MN}$ and
secondly, the tensor fields in the Lagrangian will always appear multiplied by
$Z^{MN}$. The identities proven in the previous section are essential in what
follows. 

To see how this works we first consider the gauge transformations of the
vector fields,
\begin{equation}
  \label{eq:A-var}
\delta A_\mu{}^M = \partial_\mu\Lambda^M - g\,
X_{[PQ]}{}^M\,\Lambda^P\,A_\mu{}^Q - g\,Z^{MN}\,\Xi_{\m\,N}\,,  
\end{equation}
where $\Lambda^M$ and $\Xi_{\m\,M}$ denote the parameters of the vector and
tensor gauge transformations. Observe that $X_{[MN]}{}^P$ play the role of 
generalized structure constants of the gauge group. Obviously, a number of
vector fields can be set to zero by a gauge choice. This number $t$ equals the 
rank of the matrix $Z^{MN}$. As was explained in the previous section, we
refrain from doing this in order to remain independent of the specific choice
for the embedding tensor. 

Because the Jacobi identity does not hold for the $X_{[MN]}{}^P$, the
would-be covariant field strength,
\be 
{\cal F}_{\m\n}{}^M= \pa_\m A_\n{}^M -\pa_\n A_\m{}^M + g\, X_{[NP]}{}^M
\,A_\m{}^N A_\n{}^P \,,
\ee
does not transform covariantly, 
\bea
\label{F-var}
\d{\cal F}_{\m\n}{}^M &=&{} - g\, X_{[NP]}{}^M\,\Lambda^N \,
{\cal F}_{\m\n}{}^P 
- g^2\,Z^{MN}\, d_{NP[R}\, X_{ST]}{}^P\, \Lambda^R\, A_\m{}^S\, A_\n{}^T \nn\\
&&{}{} -g\,Z^{MN} \, \Big(2\,\pa_{[\m} \Xi_{\n]N} - g\, X_{PN}{}^Q
\,A_{[\m}{}^P\,\Xi_{\nu]Q} \Big)
\nonumber \\[1.5ex]
&=&{}- g\, X_{NP}{}^M\,\Lambda^N \,{\cal F}_{\m\n}{}^P \nonumber \\
&&{}{} + g\,Z^{MN} \Big( 2\,d_{NPQ}\, \pa_{[\m} A_{\n]}{}^P
-  g\, X_{RN}{}^P \,d_{PQS}\, A_{[\mu}{}^R\,A_{\nu]}{}^S  \Big)\Lambda^Q   
 \nonumber\\
&&{}{} -g\,Z^{MN} \, \Big(2\,\pa_{[\m} \Xi_{\n]N} - g\, X_{PN}{}^Q
\,A_{[\m}{}^P\,\Xi_{\nu]Q} \Big)\,.
\eea
Up to terms proportional to $Z^{MN}$ the field strengths ${\cal F}^M$
transform covariantly under the gauge group. To regain full covariance, we
introduce tensor fields $B_{\m\n\,M}$ such that the modified field strengths, 
\be
\label{def-H}
{\cal H}_{\m\n}{}^M = {\cal F}_{\m\n}{}^M + g\, Z^{MN} \,B_{\m\n\,N}\,,
\ee
transform covariantly under the gauge group, 
\be 
\d{\cal H}_{\m\n}{}^M = - g\, X_{PN}{}^M \,\Lambda^P\,{\cal H}_{\m\n}{}^N \,,
\ee
and are invariant under the tensor gauge transformations. This implies that
the transformations of the fields $B_{\m\n\,M}$ are as follows,
\begin{eqnarray}
  \label{eq:delta-B}
  Z^{MN}\,\delta B_{\m\n\,N} &=&{} Z^{MN} \, \Big(2\,\pa_{[\m} \Xi_{\n]N} - g\,
X_{PN}{}^Q \,A_{[\m}{}^P\,\Xi_{\nu]Q} \Big)
+ g\, Z^{MN}\, \Lambda^P\,X_{PN}{}^Q \,B_{\mu\nu\,Q}   \nn\\
&&{}
{} -Z^{MN} \Big( 2\,d_{NPQ}\, \pa_{[\m} A_{\n]}{}^P
-  g\, X_{RN}{}^P \,d_{PQS} A_{[\mu}{}^R\,A_{\nu]}{}^S  \Big)\Lambda^Q   
\,. 
\end{eqnarray}
Obviously, the symmetry transformations on the tensor fields are only
determined modulo terms that vanish under contraction with $Z^{MN}$; this
poses no problem as, in the Lagrangian, the tensor fields $B_{\m\n\,M}$ will
always be contracted with $Z^{MN}$.\footnote{Note that the term $\partial_\mu \Xi_{\nu\,N}$ 
in (\ref{eq:delta-B}) is not properly covariantized by the 
next term proportional to $X_{PN}{}^Q$;  this would require an additional
factor~$2$.}

It is important to note that the covariant derivative, $D_\m= \pa_\m -g\,
A_\m{}^M\, X_M$, does not transform under tensor gauge transformations, by
virtue of \eqn{eq:Z-X-orthogonal}. Furthermore, we note the validity of the
Ricci identity,
\begin{equation}
  \label{eq:Ricci}
  {[}D_\m,D_\n] = -g\,{\cal F}_{\m\n}{}^M \,X_M\,.
\end{equation}
To verify the consistency of the above transformation rules, one may consider  
the commutator algebra of the vector and tensor gauge transformations. The
tensor transformations commute, 
\be
{[}\d({\Xi_1}), \d({\Xi_2})] =0\,.
\ee
The commutator of a vector and a tensor gauge transformation gives rise to a
tensor gauge transformation,
\be
{[}\d({\Xi}), \d({\Lambda})] = \d(\tilde\Xi) \,,
\ee
with $\tilde\Xi_{\m\,M}= \ft12
g\,X_{PM}{}^N\,\Lambda^P\,\Xi_{\m\,N}$. Finally, the commutator of two vector
gauge transformations gives rise to a vector gauge transformation and a tensor
gauge transformation,
\be
{[}\d({\Lambda_1}), \d({\Lambda_2})] = \d(\tilde\Lambda)+ \d(\tilde\Xi) \,, 
\ee
where $\tilde \Lambda^M = g\, X_{[NP]}{}^M\, \Lambda_1{}^N\,\Lambda_2{}^P$ and 
$\tilde\Xi_{\m\,M}
=-g\,d_{MN[P}\,X_{QR]}{}^N\,\Lambda_1{}^P\Lambda_2{}^Q\,A_\m{}^R$. 

There exists a kinetic term for the tensor fields which is of first-order in
derivatives, which is modified by Chern-Simons-like terms in order to be fully
gauge invariant under the combined vector and tensor gauge transformations. It
reads as follows,
\begin{eqnarray}
\label{eq:CS-term}
{\cal L}_{\rm VT} 
&\!=\!&\ft12 i \varepsilon^{\m\n\rho\sigma\tau}\Big\{ gZ^{MN}
 B_{\m\n\,M} \Big[ D_\rho B_{\sigma\tau\,N} +4\, d_{NPQ} \,A_\rho{}^P
 \Big(\partial_\sigma A_{\tau}{}^Q +\ft13 g\, X_{[RS]}{}^Q\, A_\sigma{}^R\,
 A_{\tau}{}^S\Big)\Big] \nn\\ [.5ex]
&&{}{\hspace{11mm}} - \ft83d_{MNP}\Big[ A_\mu{}^M\,\partial_\nu 
   A_\rho{}^N \,\partial_\sigma A_\tau{}^P \\
&&{}{\hspace{23mm}}+ \ft34 g\, X_{[QR]}{}^M \, A_\m{}^N A_\n{}^Q A_{\rho}{}^R
 \Big(\partial_\sigma  A_{\tau}{}^P+\ft15g\, 
 X_{[ST]}{}^P A_\sigma{}^SA_\tau{}^T \Big)\Big] \Big\}\,,
 \nn
\end{eqnarray}
where 
\be 
D_\m B_{\n\rho\,M} =
\partial_\m B_{\n\rho\,M}-g\,  X_{PM}{}^N\,A_\m{}^P B_{\n\rho\,N} \,.
\ee
Under variations $A_\m{}^M\to A_\m{}^M+ \delta A_\m{}^M$ and $B_{\mu\nu\,M}\to
B_{\mu\nu\,M}+ \delta B_{\mu\nu\,M}$ this Lagrangian changes as follows,
\begin{eqnarray}
  \label{eq:delta-Lvt}
  \delta{\cal L}_{\rm VT} &=& i\varepsilon^{\mu\nu\rho\sigma\tau} \Big\{ 
  \Big(\delta B_{\mu\nu\,M} +2\, d_{MPQ}\,\delta A_\mu{}^P \, A_\nu{}^Q \Big)
  D_\rho {\cal H}_{\sigma\tau}{}^M -
  \delta A_\mu{}^M\,d_{MNP}\,{\cal H}_{\nu\rho}{}^N\, {\cal
  H}_{\sigma\tau}{}^P \Big\}  \nn\\[1ex]
  &&{} + \mbox{ total derivatives }\cdots \,, 
\end{eqnarray}
where we note the identity,
\begin{eqnarray}
\lefteqn
   {\varepsilon^{\mu\nu\rho\sigma\tau} \,D_\mu  {\cal H}_{\nu\rho}{}^M =}  \nn\\[.5ex]
   &&{}
   \;\;
g\,\varepsilon^{\mu\nu\rho\sigma\tau} \,Z^{MN} \Big[D_\mu  B_{\nu\rho\,N} +2
\, d_{NPQ} \,A_\mu{}^P 
 \Big(\partial_\nu A_{\rho}{}^Q +\ft13 g\, X_{[RS]}{}^Q\, A_\nu{}^R
 A_{\rho}{}^S  \Big)\Big] \,. 
 \nn
\end{eqnarray}
Clearly the terms   $\varepsilon^{\mu\nu\rho\sigma\tau}\, D_\rho {\cal
  H}_{\sigma\tau}{}^M$ and
  $\varepsilon^{\mu\nu\rho\sigma\tau}\,d_{MNP}\,{\cal H}_{\nu\rho}{}^N\,
  {\cal H}_{\sigma\tau}{}^P$ will appear as part of the equations of motion
  for the vector and tensor fields. 

As we shall see later there is a second term in the supergravity Lagrangian,
quadratic in ${\cal H}_{\mu\nu}{}^M$. It requires an \Exc6-covariant metric
${\cal M}_{MN}$ that will also be discussed in due course. Here we just note
that the variation of this term in the Lagrangian under changes of the vector
and tensor fields yields, 
\begin{eqnarray}
  \label{eq:H-H-term}
  \delta({\cal M}_{MN}\,{\cal H}_{\mu\nu}{}^M \,{\cal H}^{\mu\nu\,N}) &=&
  - 2 \,\Big(\delta B_{\mu\nu\,M} -
  2\, d_{MPQ}\,A_\mu{}^P \, \delta A_\nu{}^Q \Big) Z^{MN} 
  {\cal M}_{NR} \,{\cal H}^{\mu\nu\,R} 
  \nn\\
  &&{}
  +4\,\delta A_\mu{}^M\, D_\nu\Big({\cal M}_{MN} \,{\cal H}^{\mu\nu\,N}\Big)\,,
\end{eqnarray}
which shows the same combinations of field variations as in
(\ref{eq:delta-Lvt}) so that the two variations can be combined without
difficulty.  

At this point one has the option of removing $t$ of the vector fields by
a gauge choice. To do this one employs the special \Exc6 basis of
subsection~\ref{specialbasis}. We will return to this in
subsection~\ref{gaugefixing}, where we will also exhibit the consequences of
this gauge choice for the supersymmetry transformations. In the absence of
a gauging the embedding tensor (or, equivalently, the gauge coupling constant 
$g$) vanishes, ${\cal H}_{\mu\nu}{}^M$ coincides with the abelian field
strengths ${\cal F}_{\mu\nu}{}^M$ and ${\cal L}_{\rm VT}$ reduces to an
abelian Chern-Simons term,
\begin{equation}
  \label{eq:limit-CS}
{\cal L}_{\rm VT} \longrightarrow - \ft43 i
 \varepsilon^{\m\n\rho\sigma\tau}\, d_{MNP}\, A_\mu{}^M\,\partial_\nu 
   A_\rho{}^N \,\partial_\sigma A_\tau{}^P \,. 
\end{equation}

\section{${\rm E}_{6(6)}/{\rm USp}(8)$ and the $T$-tensor}
\setcounter{equation}{0}
\label{T-tensor}
We already stressed in the introduction that the scalar fields parametrize the
${\rm E}_{6(6)}/{\rm USp}(8)$ coset space. These fields are described by a
matrix ${\cal V}(x) \in {\rm E}_{6(6)}$ (taken in the fundamental ${\bf 27}$
representation) which transforms from the right under local ${\rm USp}(8)$ and
from the left under rigid \Exc6. The matrix $\vv$ can be used to elevate the
embedding tensor to the so-called $T$-tensor, which is the 
${\rm USp}(8)$-covariant, field-dependent, 
tensor that appears in the masslike terms and the scalar potential. The
$T$-tensor is thus defined by,
\be
T_{\underline M}{}^{\underline{\a}}[\Theta,\phi]\,t_{\underline\a}  =
{\cal V}^{-1}{}_{\!\!\!\!\underline{M}}{}^N\, 
\Theta_N{}^\a\,  ({\cal V}^{-1}t_\a {\cal V} )\;,
\ee
where the underlined indices refer to local ${\rm USp}(8)$. The appropriate
representation is the ${\bf 27}$, so that we can write
\begin{equation}
  \label{eq:T-theta}
T_{\underline M \underline N}{}^{\underline P}[\Theta,\phi]  =
{\cal V}^{-1}{}_{\!\!\!\!\underline{M}}{}^M\, {\cal
  V}^{-1}{}_{\!\!\!\!\underline{N}}{}^N\, {\cal
  V}_{{P}}{}^{\underline  P}\, X_{MN}{}^P\,.
\end{equation}
When treating the embedding tensor as a spurionic object that
transforms under the duality group, the Lagrangian and
transformation rules remain formally invariant under \Exc6. Under
such a transformation $\Theta$ would transform as 
$\Theta_M{}^\a\, t_\a \to g_M{}^N\,\Theta_N{}^\a\,(g\,t_\a g^{-1})$, 
with $g \in {\rm E}_{6(6)}$.  Of
course, when freezing $\Theta_M{}^\a$ to a constant, the \Exc6-invariance is
broken. 

It is clear that the \Exc6-covariant constraints on the embedding tensor are
in direct correspondence to a set of ${\rm USp}(8)$-covariant constraints on
the $T$-tensor, as their assignment to representations of \Exc6 and/or its
subgroups are directly related. To make this more explicit we note that 
{\it every} variation of the coset representative can 
be expressed as a (possibly field-dependent) \Exc6 transformation
acting on ${\cal V}$ from the right. For example, a rigid \Exc6 
transformation acting from the left can be rewritten as a field-dependent
transformation from the right, 
\be
{\cal V}\to {\cal V}^\prime = g\,{\cal V}= {\cal V} \, \sigma^{-1}\,,
\ee
with $\sigma^{-1}= {\cal V}^{-1}\,g\,{\cal V}\in {\rm E}_{6(6)}$, but also a 
supersymmetry transformation can be written in this
form. Consequently, these variations of ${\cal V}$ induce the following
transformation of the $T$-tensor,
\be
\label{T-linear}
T_{\underline M \underline N}{}^{\underline P}\, 
\to T^{\prime}_{\underline M\underline N}{}^{\underline P} =
\sigma_{\underline M}{}^{\underline Q}\,\sigma_{\underline N}{}^{\underline R}
\,(\sigma^{-1})_{\underline S}{}^{\underline P}
\,T_{\underline Q \underline R}{}^{\underline S} \;. 
\ee
This implies that the $T$-tensor constitutes a representation of
\Exc6. Observe that this is {\it not} an invariance statement; rather it means
that the $T$-tensor (irrespective of the choice for the corresponding
embedding tensor) varies under supersymmetry or any other transformation in a
way that can be written as a (possibly field-dependent)
\Exc6-transformation. Note also that the transformation assignment of 
the embedding tensor and the $T$-tensor are opposite in view of the
relationship between $g$ and $\sigma$, something that is important in
practical applications.  

The maximal compact ${\rm  USp}(8)$ subgroup of \Exc6 coincides with the
R-symmetry group that acts on the fermion fields: the gravitino and spinor
fields are symplectic Majorana spinors transforming in the ${\bf 8}$, and 
${\bf 48}$ representation, respectively. A crucial role is played here by the 
${\rm USp}(8)$-invariant skew-symmetric tensors $\Omega_{AB}$ and 
$\Omega^{AB}=(\Omega_{AB})^\ast$  ($A,B,\ldots= 1,\ldots,8$) satisfying 
$\Omega^{AC}\,\Omega_{CB}  =-\d^A{}_B$. 
The presence of the representations of the spinor fields leads one to adopt
a notation for \Exc6 vectors $x_M$ and $y^M$
where indices $M$ are replaced by antisymmetric, symplectically traceless,
index pairs $[AB]$. 
The ${\bf 27}$ representation is thus described by a pseudoreal, 
  antisymmetric and symplectic traceless tensor $x_{AB}$,
\begin{eqnarray}
  \label{eq:E6-basis}
  x^{AB} \equiv (x_{AB})^\ast = \Omega^{AC}\,\Omega^{BD}\,x_{CD}\,, \qquad
  \Omega^{AB}\,x_{AB}=0 \,.
\end{eqnarray}
Raising and lowering of indices is effected by complex
conjugation. Corresponding identities hold for the $y^{AB}$ transforming in
the $\overline{\bf 27}$ representation. The action of infinitesimal \Exc6
transformations reads as follows,
\begin{eqnarray}
  \label{eq:delta-x-y}
\d x_{AB} &=& -2\, \Lambda_{[A}{}^C \,x_{B]C}+ \Sigma_{ABCD} \,x^{CD}\,,\nn\\
\d y^{AB} &=&  2\,\Lambda_C{}^{[A} \,y^{B]C}- \Sigma^{ABCD} \, y_{CD}\,,
\end{eqnarray}
so that $x_{AB}\,y^{AB}$ is an \Exc6 invariant. 
Here $\Lambda_A{}^B$ parametrizes the ${\rm USp}(8)$ transformations; the 
fully antisymmetric pseudoreal and symplectic traceless tensors $\Sigma$
parametrize the remaining \Exc6 transformations in accord with the following
decomposition of the adjoint representation of \Exc6: 
${\bf 78}\to {\bf 36} + {\bf 42}$. The explicit 
restrictions on the parameters read,
\bea
  \label{eq:E6}
&&{}\Lambda^A{}_B \equiv (\Lambda_A{}^B )^\ast =   -\Lambda_B{}^A\,,\quad
\Lambda_{[A}{}^C\,\Omega_{B]C}=0\,, \quad \Lambda_A{}^A = 0\,, 
\quad \Sigma_{ABCD}=\Sigma_{[ABCD]} \,, 
\nn\\
&&{}
\Sigma_{ABCD} \equiv (\Sigma^{ABCD})^\ast =
\Omega_{AE}\,\Omega_{BF}\,\Omega_{CG}\,\Omega_{DH}\,\Sigma^{EFGH}\,,\quad
\Omega^{AB} \,\Sigma_{ABCD}=0\,. 
\end{eqnarray}
Furthermore we note the identity, 
\begin{equation}
  \label{eq:irreducible}
  \Omega_{[AB}\,\Sigma_{CDEF]}= 0\,,
\end{equation}
which holds by virtue of the fact that $\Sigma$ transforms in the irreducible
${\bf 42}$ representation of ${\rm USp}(8)$. From it one derives that
$\Omega_{AB}\,y^{BC}\,\Omega_{CD}\,y^{DE}\,\Omega_{EF}\,y^{FA}$ is an \Exc6
invariant and thus represents the invariant symmetric tensor $d_{MNP}$
that we encountered in previous sections. From \eqn{eq:irreducible} we also
derive the identities,  
\begin{eqnarray}
  \label{eq:sigma-algebra}
&& \Sigma_{1\,ACDE} \,\Sigma_2{}^{BCDE} + \Sigma_{2\,ACDE} \,\Sigma_1{}^{BCDE}
 =\ft1{16}\,\d_A{}^B \,\Sigma_{1\,CDEF} \,\Sigma_2{}^{CDEF}\,, \nonumber\\[.5ex]
&& \Sigma_{1\,ABEF} \,\Sigma_2{}^{EFCD} \,- \Sigma_{2\,ABEF} \,\Sigma_1{}^{EFCD}=
\nn\\ 
&&\qquad\qquad\qquad\qquad\qquad
\ft23\,\d_{[A}{}^{[C} \,(\Sigma_{1\,B]EFG}
\,\Sigma_2{}^{D]EFG}-\Sigma_{2\,B]EFG} \,\Sigma_1{}^{D]EFG}) \,.
\end{eqnarray}
The term on the right-hand side of the second equation parametrizes an
infinitesimal ${\rm USp}(8)$ transformation. This identity ensures the closure
of the algebra associated with \Exc6. 

In supergravity, which we will discuss in the next section, we will use a
hybrid notation where rigid \Exc6 indices will be denoted by $M,N,\ldots$,
whereas the {\it local} ${\rm USp}(8)$ indices $\underline M,\underline
N,\ldots$ will be replaced by antisymmetric, symplectic traceless, index pairs
$[ij]$. The indices $i,j,\ldots$ are also carried by the fermion fields; those
fields carry an odd number of indices\footnote{
  Note that only even-rank tensors can be pseudoreal, so that fermions
  transform in complex representations of ${\rm USp}(8)$. Observe also that
  $\Omega$ is pseudoreal, {\it i.e.},   $\Omega_{ij} =
  \Omega_{ik}\,\Omega_{jl}\,\Omega^{kl}$. }. 
With these conventions the
coset representative ${\cal V}_M{}^{\underline N}$ is  
written as ${\cal V}_{M}{}^{ij}$, with ${\cal
  V}_{M}{}^{ij}\,\Omega_{ij}=0$. Furthermore ${\cal V}_{M}{}^{ij}$ is 
pseudoreal: ${\cal V}_{M\,ij}\equiv 
({\cal V}_{M}{}^{ij})^\ast = {\cal V}_{M}{}^{kl}\,\Omega_{ki}
\,\Omega_{lj}$. Denoting the inverse of ${\cal V}_{M}{}^{ij}$ by 
${\cal V}_{ij}{}^M$, we have\footnote{
  Observe that the double counting associated with summing over antisymmetric
  index pairs implies a change of the inner product as the direct conversion
  $X_{\underline M}\to X_{ij}$ and $Y^{\underline M}\to Y^{ij}$ yields
  $X_{\underline M}Y^{\underline M}= 2\,\sum_{i>j} 
  X_{ij}Y^{ij}$. The factor 2 can in principle be absorbed into the
  definitions  of $X_{ij}$ and $Y^{ij}$, but we refrain from doing so.},
\begin{eqnarray}
  \label{eq:inverse}
  {\cal V}_{M}{}^{ij}\,{\cal V}_{ij}{}^{N} &=& \d_{M}{}^{N}\,,\nn\\
  {\cal V}_{ij}{}^{M}\, {\cal V}_{M}{}^{kl} &=& \d_{ij}{}^{kl} 
  -\ft18 \Omega_{ij}\,\Omega^{kl}\,.
\end{eqnarray}

Exploiting the group property it follows that any variation $\Delta\vv$
takes the form, 
\begin{eqnarray}
  \label{eq:P-Q}
  \Delta \vv_{M}{}^{ij} - \vv_{M}{}^{kl} \Big(2\, \d_k{}^{[i}\,{\cal
  Q}_{l}{}^{j]} +{\cal P}^{ijmn}\,\Omega_{mk}\,\Omega_{nl} \Big) =0  \,,
\end{eqnarray}
where ${\cal Q}$ and ${\cal P}$ span the \Exc6 algebra, so that they transform
according to the ${\bf 36}$ and ${\bf 42}$ representation of ${\rm USp}(8)$,
respectively. Consequently ${\cal Q}$ and ${\cal P}$ are subject to the
conditions \eqn{eq:E6} (but 
with local ${\rm USp}(8)$ rather than  with rigid \Exc6 indices). It is
straightforward to find explicit expressions for ${\cal Q}$ and ${\cal P}$ in
terms of $\vv^{-1}\Delta\vv$,
\begin{eqnarray}
  \label{eq:PQ-solutions}
 {\cal Q}_i{}^j&=& \ft13 \,\vv_{ik}{}^{M} \,\Delta\vv_{M}{}^{jk}\,,
 \nn\\ 
 {\cal P}^{ijkl}&=& \vv_{mn}{}^{M}
 \,\Delta\vv_{M}{}^{[ij}\,\Omega^{k\vert m\vert}\,\Omega^{l]n} \,,
\end{eqnarray}
where $\vert m\vert$ indicates that the index $m$ is exempted from the
antisymmetrization which pertains to $[ijkl]$. 
Possible variations $\Delta\vv$ include spacetime
derivatives $\pa_\m\vv$ or a gauge transformation with parameters
$\Lambda^M$ according to 
$\Delta \vv_{M}{}^{ij} = \Lambda^{P}\,X_{PM}{}^{N} \,\vv_{N}{}^{ij}$. 
When $\Delta$ denotes the gauge-covariant derivative $\pa_\m -
g\,A_\m{}^{M}\,X_{M}$, \eqn{eq:P-Q} defines the ${\rm USp}(8)$ composite
connection ${\cal Q}_{\mu\,i}{}^j$ in the presence of the gauging, while 
${\cal P}_\m{}^{ijkl}$ is the ${\rm USp}(8)$-covariant tensor whose square 
will constitute the kinetic term for the scalar fields,
\begin{eqnarray}
  \label{eq:PQ-mu}
 {\cal Q}_{\m\,i}{}^j&=& \ft13 \,\vv_{ik}{}^{M} \,\pa_\mu
 \vv_{M}{}^{jk}  - g \,  A_\m{}^{M} \,Q_{M\,i}{}^j\,, \nn\\ 
 {\cal P}_\m{}^{ijkl}&=&  \vv_{mn}{}^{M}
 \,\pa_\mu\vv_{M}{}^{[ij}\,\Omega^{k\vert m\vert}\,\Omega^{l]n} 
- g\,A_\mu{}^{M} \, {\cal P}_{M}{}^{ijkl} \,.
\end{eqnarray}
Here we used the definitions, 
\begin{eqnarray}
  \label{eq:PQ-AB}
 {\cal Q}_{M\,i}{}^j&=& \ft13 \,\vv_{ik}{}^{N}
 \,X_{MN}{}^{P}\vv_{P}{}^{jk}\,, \nn\\ 
 {\cal P}_{M}{}^{ijkl}&=& \vv_{mn}{}^{N}
 \,X_{MN}{}^{P}\vv_{P}{}^{[ij}\,\Omega^{k\vert m\vert}\,\Omega^{l]n} \,. 
\end{eqnarray}
We note that ${\cal P}_{M}{}^{ijkl}$ represents the Killing vectors of the
  ${\rm E}_{6(6)}/{\rm USp}(8)$ coset space associated with its gauged
  isometries.

Here and henceforth we will use derivatives $D_\m$ that are covariant with
respect both local ${\rm USp}(8)$ (with composite connections ${\cal
  Q}_{\m\,i}{}^j$) and the subgroup of \Exc6 that is gauged by (a subset
of) the vectors $A_\m{}^{M}$. Applying two such derivatives on $\vv$ leads to
an integrability relation upon antisymmetrization, 
\begin{eqnarray}
  \label{eq:CM-eq}
F_{\mu\nu}({\cal Q})_i{}^j & \equiv & \pa_\m {\cal Q}_{\n\,i}{}^j - \pa_\n
  {\cal   Q}_{\m \,i}{}^j + 2\,{\cal Q}_{[\m\,i}{}^k {\cal Q}_{\n]\,k}{}^j
  \nn\\ 
 &=& 
  - \ft23\,{\cal P}_{[\mu\,iklm}\, {\cal P}_{\nu]}{}^{j klm} - g\,
  {\cal F}_{\m\n}{}^{M} \,{\cal Q}_{M\,i}{}^j\,, 
\nonumber \\[.5ex]
  D_{[ \mu} {\cal P}_{\nu]}{}^{ijkl}  & = & - \ft12 g\,{\cal F}_{\m\n}{}^{M} \,
  {\cal P}_{M}{}^{ijkl}  \,,
\end{eqnarray}
where we made use of \eqn{eq:Ricci}. These are the Cartan-Maurer equations
with extra terms of order $g$ induced by the gauging. In the Lagrangian those
terms initially cause a  breaking of supersymmetry that will have to be
compensated by new interaction terms and variations.  The order-$g$
corrections in \eqn{eq:CM-eq} are in fact proportional to the two components
of the $T$-tensor defined already in (\ref{eq:T-theta}),
\begin{eqnarray}
  \label{eq:T-tensor}
  T^i{}_{jmn}&=&  {\cal Q}_{M\,j}{}^i\, \vv_{mn}{}^{M}\,, \nn\\
  T^{ijkl}{}_{mn}&=& {\cal P}_{M}{}^{ijkl}\,\vv_{mn}{}^{M}\,.
\end{eqnarray}
Both these components are pseudoreal and have symmetry properties that should
be obvious from the preceding text. In particular, note that 
$T_i{}^{jkl}= \Omega_{im}\,\Omega^{jn}\,\Omega^{kp}\,\Omega^{lq}
\,T^m{}_{npq}$ and 
$\Omega_{k[i} \,T^k{}_{j]mn} = \Omega^{k[i}\, T^{j]}{}_{kmn}=0$, so that
$T_i{}^{jkl} = - T^j{}_{imn}\,\Omega^{mk}\, \Omega^{nl}$. 
We note the following convenient relation,
\begin{equation}
  \label{eq:def-X-T}
  X_{MN}{}^P= \vv_M{}^{mn}\,\vv_N{}^{kl}\,\vv_{ij}{}^P \Big[
  2\,\delta_k{}^i\,T^j{}_{lmn} +
  T^{ijpq}{}_{mn}\,\Omega_{pk}\,\Omega_{ql}\Big]   \,, 
\end{equation}
which is just the inverse of (\ref{eq:T-theta}).

Following the argument given at the beginning of the section, we consider 
variations of the coset representative. These can always be cast in the form
of a \Exc6 transformation acting on the 
right of $\vv$, which implies that any variation of the $T$-tensor is again
proportional to the $T$-tensor itself. Since the variations under ${\rm
  USp}(8)$ are obvious, the relevant variation concerns 
\begin{equation}
  \label{eq:V-variation}
\delta \vv_{M}{}^{ij} = - \vv_{M}{}^{mn}\,\Omega_{mk}\,
  \Omega_{nl}\, \Sigma^{ijkl}\,. 
\end{equation}
It is straightforward to determine the effect of this variation on various
${\rm USp}(8)$ tensors and we present the explicit results, 
\begin{eqnarray}
  \label{eq:sigma-transformations}
  \d{\cal Q}_{\m\,i}{}^j &=&- \ft13 {\cal P}_{\mu\,iklm}\,\Sigma^{jklm} 
   + \ft13 \Sigma_{iklm}\, {\cal P}_\m{}^{jklm}  \,, \nn\\
  \d{\cal P}_\m{}^{ijkl}&=& - D_\mu \Sigma^{ijkl}\,, \nn\\
  \d T^i{}_{jmn}&=& \ft13 \Sigma_{jpqr}\, T^{ipqr}{}_{mn} 
  - \ft13 \Omega^{iv}\,\Omega_{jw} \,\Sigma_{vpqr} \,T^{wpqr}{}_{mn} +
  \Sigma_{mnpq} \,\Omega^{pr}\,\Omega^{qs}  \,T^i{}_{jrs} \,,\nn\\
  \d T^{ijkl}{}_{mn}&=& - 4\,  T^{[i}{}_{p mn}\,\Sigma^{jkl]p} + \Sigma_{mnpq}
  \,\Omega^{pr} \,\Omega^{qs}  \,
  T^{ijkl}{}_{rs} \,.
\end{eqnarray}

Armed with these results we can now proceed and derive the constraints on the
$T$-tensor that are induced by the corresponding constraints on embedding
tensor discussed in section~2. First of all, the $T$-tensor will be
restricted as a result of the representation constraint
\eqn{eq:repres-constraint}, according to which the embedding tensor belongs to
the  ${\bf 351}$ representation of \Exc6. This representation branches
under ${\rm USp}(8)$ into ${\bf 36} +{\bf 315}$. According to
\eqn{eq:sigma-transformations} the $T$-tensor should therefore precisely
comprise these two representations. Here it  
is helpful to indicate the ${\rm USp}(8)$ representations that are described
by the unconstrained $T$-tensor,
\begin{eqnarray}\label{eq:1}
  \label{eq:T-tensor-decom}
  T^i{}_{jkl} &:& {\bf 36}\times {\bf 27}= \underline{\bf 36}+ 
\underline{\bf 315}+  {\bf 27}   +{\bf 594}\,, \nn\\
 T^{ijkl}{}_{mn} &:&    {\bf 42}\times {\bf 27}= \underline{\bf 315}
+ {\bf 27}+  {\bf 792}\,,
\end{eqnarray}
where the underlined representations are those that are allowed by the
constraint. Therefore, when subject to the constraint, $T^{klmn}{}_{ij}$ will
exclusively belong to 
the ${\bf 315}$ representation, while $T^{i}{}_{jlm}$ is decomposable into
the ${\bf 315}$ and the ${\bf 36}$ representations. Describing these two
representations by two pseudoreal, symplectic traceless, tensors $A_1^{ij}$
and $A_2{}^{i,jkl}$, satisfying 
$A_1^{[ij]}=0$, $A_{2}{}^{i,jkl} = A_{2}{}^{i,[jkl]}$ and 
$A_{2}{}^{[i,jkl]}=0$,
one can write down the following decomposition of the $T$-tensor,
\begin{eqnarray}
  \label{eq:usp8-repre-constraint}
T^{klmn}{}_{ij} &=& 4\, A_2{}^{q,[klm}\, \d^{n]}{}_{[i}\, 
\Omega_{j]q} + 3\, A_2{}^{p,q[kl}\,\Omega^{mn]}\, 
\Omega_{p[i}\,\Omega_{j]q}  \,, \nonumber \\   
T_i{}^{jkl} &=& -\Omega_{im}\,A_2{}^{(m,j)kl} - \Omega_{im} \Big(
\Omega^{m[k} \,A_1{}^{l]j} +\Omega^{j[k} \,A_1{}^{l]m}+ \ft14
\Omega^{kl}\,A_1{}^{mj} \Big) \,,
  \end{eqnarray}
where the relative factor on the right-hand side (overall factors can be
absorbed into $A_{1,2}$) are determined by the fact that $A_1$ and $A_2$
together constitute an irreducible representation of \Exc6. Algebraically, the
factor follows from requiring consistency of
(\ref{eq:usp8-repre-constraint})with the transformation rules 
(\ref{eq:sigma-transformations}). Explicit evaluation, making repeated use of 
(\ref{eq:irreducible}), leads to the following variations for the tensors
$A_1$ and $A_2$, 
\begin{eqnarray}
  \label{eq:A-variations}
    \d A_1{}^{ij} &=& \ft49 \,\Omega^{p(i}\,\Sigma^{j)klm}\, A_{2\,p,klm}
    \,,\nn\\  
    \d A_2{}^{i,jkl} &=& \ft32\,\Big(\Omega^{mi}\, \Sigma^{jkln}
    + \Omega^{m[j}\, \Sigma^{kl]in} \Big) \,A_{1\,mn}\nn\\
    &&{} -\Big(\Omega^{i[j}\, \Omega^{k|m|}\, \Sigma^{l]npq }
       -3  \,\Omega^{ni}\, \Omega^{m[j}\, \Sigma^{kl]pq} \nn\\
    &&{} \hspace{6mm}- \ft16 \Omega^{im}\,
    \Omega^{[kl} \, \Sigma^{j]npq} +\ft16 \Omega^{m[j}\,\Omega^{kl]}
    \,\Sigma^{inpq} \Big) \, A_{2\,m,npq} \,.
\end{eqnarray}
We note that these variations are consistent with the ${{\rm USp}(8)}$
irreducibility constraints for the tensors $A_{1,2}$ themselves. Furthermore   
we note that the linear combination 
\begin{equation}
  \label{eq:usp8-z}
  {\cal Z}^{ij,kl} \equiv \Omega^{[i[k}\,A_1{}^{l]j]} + A_2{}^{[i,j]kl} \,,
\end{equation}
defines an antisymmetric tensor in the symplectic traceless index pairs $[ij]$
and $[kl]$, transforming according to 
\begin{equation}
  \label{eq:delta-z}
   \delta {\cal Z}^{ij,kl} =- \Sigma^{ijmn}\,\Omega_{mp}\,\Omega_{nq} \,
   {\cal Z}^{pq,kl} - \Sigma^{klmn}\,\Omega_{mp}\,\Omega_{nq} \, 
   {\cal Z}^{ij,pq}  \,.  
\end{equation}
This shows (c.f. (\ref{eq:delta-x-y})) that ${\cal Z}^{ij,kl}$ must be the
dressed version of the \Exc6 tensor $Z^{MN}$,
\begin{equation}
  \label{eq:ZZ}
  {\cal Z}^{ij,kl} =\ft15 \sqrt{5}\, \vv_M{}^{ij}\, \vv_N{}^{kl}\, Z^{MN}\,. 
\end{equation}
The proportionality constant follows from applying (\ref{eq:a-sym-X}),
with $X_{MN}{}^P$ expressed by (\ref{eq:def-X-T}), employing the following
representation of the invariant symmetric three-rank tensors,   
\begin{eqnarray}
  \label{eq:d-symbol}
  d_{MNP} &=& \ft25\sqrt5\, \vv_M{}^{ij} \,\vv_N{}^{kl}\, \vv_P{}^{mn}
  \,\Omega_{jk} \Omega_{lm}\,\Omega_{ni}   \,,   \nn\\ 
  d^{MNP} &=&  \ft25\sqrt5\, \vv_{ij}{}^M \,\vv_{kl}{}^N\,
  \vv_{mn}{}^P \, \Omega^{jk}  \Omega^{lm}\,\Omega^{ni} 
   \,.
\end{eqnarray}
To derive the above representation we made use of the observation below
(\ref{eq:irreducible}). Note that the constancy of $d_{MNP}$ and $d^{MNP}$ is
ensured by \Exc6 invariance and that the normalization is in accord
with~(\ref{eq:d-normalization}).  

The variations (\ref{eq:A-variations}) can be used to determine the
supersymmetry variations of these tensors (as we will discuss in the next
section). We also note the following expressions for the ${\rm USp}(8)$-covariant derivatives of $A_{1,2}$,
\begin{eqnarray}
  \label{eq:A-derivatives}
    D_\m A_1{}^{ij} &=& - \ft49 \,\Omega^{p(i}\,{\cal P}_\mu{}^{j)klm}\,
    A_{2\,p,klm}   \,,\nn\\  
    D_\m A_2{}^{i,jkl} &=& - \ft32\,\Big(\Omega^{mi}\, {\cal P}_\mu{}^{jkln}
    + \Omega^{m[j}\, {\cal P}_\mu{}^{kl]in} \Big) \,A_{1\,mn}\nn\\
    &&{} +\Big(\Omega^{i[j}\, \Omega^{k|m|}\, {\cal P}_\mu{}^{l]npq }
       -3  \,\Omega^{ni}\, \Omega^{m[j}\, {\cal P}_\mu{}^{kl]pq} \nn\\
    &&{} \hspace{6mm}- \ft16 \Omega^{im}\,
    \Omega^{[kl} \, {\cal P}_\mu{}^{j]npq} +\ft16 \Omega^{m[j}\,\Omega^{kl]}
    \,{\cal P}_\mu{}^{inpq} \Big) \, A_{2\,m,npq} \,.
\end{eqnarray}

Having determined the consequences of the representation constraint, it
remains to derive the consequences of the closure constraint
(\ref{eq:closure-constraint}). This will lead to identities quadratic in the
$T$-tensor. Here we have the option of using either one of the equivalent
version presented in (\ref{eq:equiv-quadr}). It is convenient to choose the
second one and write it in terms of the $T$-tensor and the tensor
(\ref{eq:usp8-z}). The constraint then implies that the following products of
these tensors should vanish, 
\begin{equation}
  \label{eq:z-T}
   T^i{}_{jkl}\, {\cal Z}^{kl,mn} =0= T^{ijkl}{}_{mn}\,{\cal Z}^{mn,pq}\,.
\end{equation}
These two identities take the form,
\begin{eqnarray}
  \label{TZ=0}
0&=&
\delta^{[i}{}_{(k}\,A_{1}{}^{j]m}\,A_{1\,l)m} 
+  A_{2\,(k,l)mn}\,A_{1}{}^{m[i}\,\Omega^{j]n} 
\nonumber\\
&&{}{}
-2\,A_{2}{}^{[i,j]mn} \, A_{1\,m(k}\, \Omega_{l)n} 
- A_{2}{}^{m,nij}\,A_{2\,(k,l)mn}
\;,
\nonumber\\[1ex]
0&=&
- 2\, A_{2}{}^{k,[pqr}\, \delta^{s]}{}_{[i} \,\,A_{1\,j]k}
- 2\,A_{2}{}^{k,[pqr}\,\Omega^{s]l} \,\Omega_{k[i} \,A_{1\,j]l}
-3\,\Omega^{[pq}\,A_{2}{}^{r,s]kl} \,A_{1\,k[i}\,\Omega_{j]l}
\nonumber\\[1ex]&&{}{}
+4\,  A_{2}{}^{k,[pqr}\,\Omega^{s]l}\,A_{2\, [i,j]kl} 
+3\,\Omega^{[rs}\,A_{2}{}^{p,q]mn}\,A_{2\,m,nij}\;,  
\end{eqnarray}
By contraction one derives three equations, which can be written as follows,
\begin{eqnarray}
  \label{eq:A-square}
 0 &=& A_{1\,kl}\,A_2{}^{k,lmi}\,\Omega_{mj} + A_{2\,k,lmj}\,A_2{}^{k,lmi}
  -\ft19 A_{2\,j,klm}\, A_2{}^{i,klm} -\ft19 \delta^i{}_j 
 \vert A_2{}^{k,lmn}\vert^2 \,,
  \nn \\ [1ex]
 0&=& 3\,A_1{}^{im}\,A_{1\,jm} -\ft13 A_2{}^{i,klm} \,A_{2\,j,klm}
  -\ft18 \delta^i{}_j  
  (3\, \vert A_1{}^{kl}\vert^2 -\ft13 \vert A_2{}^{k,lmn}\vert^2 )\,,
  \nn\\[1ex]   
 0&=& ( \ft14 A_{1\,mn}\,\Omega_{pi} - A_{2\,[m,n]pi})
 A_2{}^{m,np[j}\,\Omega^{kl]} \nn\\
 &&{}
 + (\ft12 A_{1\,p(m} \Omega_{n)i} + A_{2\,(m,n)pi} + A_{2\,i,mnp})
 A_2{}^{m,n[jk}{} \Omega^{l]p} \nn\\
 &&{}
 + \ft13 A_{1\,im}\,A_2{}^{m,jkl} +\ft12 A_{1\,mn}\,
   A_2{}^{m,n[jk}\,\delta^{l]}{}_i  \,.
\end{eqnarray}
The first equation (\ref{TZ=0}) can be written with a similar index structure
as the last equation (\ref{eq:A-square}),
\begin{eqnarray}
  \label{eq:TZ-1}
\lefteqn{\ft13 A_{1\,im}\,A_2{}^{m,jkl} + (\ft12 A_{1\,p(m} \Omega_{n)i} +
 A_{2\,(ip)mn} )  A_2{}^{m,n[jk}{} \Omega^{l]p}} \nn\\ 
&&{}\qquad\qquad
= - \ft1{18} \Big(\delta_i{}^{[j}\,\Omega^{k\vert q\vert}  \,
 A_2{}^{l],mnp}\,A_{2\, q,mnp} - 
\Omega^{[jk}  \, A_2{}^{l],mnp}\,A_{2\, i,mnp} \Big)  \,,
\end{eqnarray}
where we also made use of the second equation (\ref{eq:A-square}). The same
applies to the first equation (\ref{eq:A-square}), which, when combined with
the third equation of (\ref{eq:A-square}) and with (\ref{eq:TZ-1}), yields an
identity that only involves terms quadratic in  $A_2$,
\begin{eqnarray}
\label{A_2-square}
 0&=& A_{2\,m,npi}\,A_2{}^{m,n[jk}\,\Omega^{l]p} + \ft12 A_{2\,m,npq}\,
 A_2{}^{m,np[j} \,\Omega^{k\vert q \vert} \, \delta^{l]}{}_i \nn\\
&&{}
 -\ft12
 (A_{2\,m,npi} \,A_2{}^{m,np [j} -\ft16 A_{2\,m,npq}
 \,A_2{}^{m,npq}\,\delta_i{}^{[j} ) \Omega^{kl]} \,. 
\end{eqnarray}
Somewhat surprisingly, this identity is not implied by (\ref{eq:z-T}) as we
have also been able to derive it directly, without making use of
(\ref{eq:z-T}). The identity  simply reflects the fact that the symmetric
product of two ${\bf 315}$  representations contains only a single ${\bf 315}$ 
representation.  

\section{Lagrangian and supersymmetry transformations}
\setcounter{equation}{0}
\label{lagrangian}
The previous results play a crucial role in establishing the supersymmetry of
the action of five-dimensional maximal gauged supergravity. The various
gaugings are encoded in the embedding tensor. When
treating the embedding tensor as a (spurionic) \Exc6-covariant tensor, the
action will be manifestly \Exc6 invariant, irrespective of the gauge group,
provided that its corresponding embedding tensor satisfies the constraints
outlined previously.  

Five-dimensional world and tangent-space indices are denoted by
$\mu,\nu,\ldots$ and $a,b,\ldots$, respectively, and take the values
$1,2,\ldots,5$. We employ hermitean $4\times4$ gamma matrices $\gamma_a$, which
satisfy 
\begin{eqnarray} 
  \label{eq:spinor-conv} 
  C\gamma_a C^{-1} &=& \gamma_a{}^{\rm T}\,,\qquad C^{\rm T} = -C\nn\,, \qquad
  C^\dagger = C^{-1} \,, \\ 
  \gamma_{abcde} &=& {\bf 1}\,\varepsilon_{abcde} \,.
\end{eqnarray}  
Here $C$ denotes the charge-conjugation matrix and gamma matrices with $k$
multiple indices denote 
the fully antisymmetrized product of $k$ gamma matrices in the usual fashion,
so that we have, for instance, $\gamma_a\,\gamma_b={\bf 1} \,\delta_{ab} +
\gamma_{ab}$. In view of the last equation, gamma matrices with more than two
multiple indices are not independent, and can be linearly expressed into the
unit matrix, $\gamma_a$ and $\gamma_{ab}$. Note that $C$, $C\gamma_a$ and 
$C\gamma_{ab}$ constitute a complete basis of 6 
antisymmetric and 10 symmetric (unitary) matrices in spinor 
space. The gamma matrices commute with the automorphism group of the Clifford
algebra, ${\rm  USp}(2N)$, where $N$ denotes the
number of independent spinors.  In the case at hand we have $N=4$.
Spinors can be described either as Dirac spinors, or as symplectic Majorana
spinors. The latter description is superior in that it makes the
action of the ${\rm USp}(8)$ R-symmetry group manifest. We will thus employ
symplectic Majorana spinors $\psi^i$ with $i=1,2,\ldots,8$, subject to the
reality constraint,    
\begin{equation}
  \label{eq:Majorana}
  C^{-1} \,\bar\psi_i {}^{\rm T}= \Omega_{ij}\,\psi^j\,,
\end{equation}
where $\Omega$ is the symplectic ${\rm USp}(8)$-invariant tensor introduced
previously. 
Observe that we adhere to our convention according to which raising or
lowering is effected by complex conjugation. 

The gravitini $\psi_\mu{}^i$ and associated supersymmetry parameters
$\epsilon^i$ transform in the ${\bf 8}$ representation of ${\rm USp}(8)$,
whereas the spinor fields $\chi^{ijk}$ transform in the ${\bf 48}$
representation. The symplectic Majorana constraint for the latter reads,
\begin{equation}
  \label{eq:chi-Majorana}
  C^{-1} \,\bar\chi_{ijk} {}^{\rm T}= \Omega_{il}\, \Omega_{jm}\,
  \Omega_{kn}\,\chi^{lmn} \,, 
\end{equation}

Finally we note the following relation for fermionic bilinears, with spinor
fields $\psi^i$ and $\varphi^i$,
\begin{equation}
  \label{eq:bilinear}
  \bar \psi_i\Gamma \varphi^j= -  \Omega_{ik}\,\Omega^{jl}\,  \bar \varphi_l
  (C^{-1}\, \Gamma^{\rm T}\,C)  \psi^k \,. 
\end{equation}
Comparing this to the hermitean conjugate of these bilinears, one finds that
$i\,\bar \psi_i\, \varphi^j$, 
$\bar \psi_i\gamma_a \varphi^j$ and 
$i\,\bar \psi_i\gamma_{ab} \varphi^j$ are pseudoreal.

Rather than first deriving the Lagrangian, we start by considering the
transformation rules, 
restricted by \Exc6, vector-tensor gauge invariance, and other invariances, up
to terms of higher order in the fermion fields. The coefficients in these
variations (apart from certain normalizations) can be fixed by requiring that
the supersymmetry closes up that order, 
\begin{eqnarray}
  \label{eq:susy-trans}
  \delta e_\mu{}^a &=&\ft12 \bar \epsilon_i\gamma^a \psi_\m{}^i \,,\nn\\
  \delta {\cal V}_{M}{}^{ij} &=&i\, {\cal V}_{M}{}^{kl}
   \Big[4\,\Omega_{p[k} \bar\chi_{lmn]} \epsilon^{p}+ 3\,\Omega_{[kl}
  \bar\chi_{mn]p} \epsilon^{p} \Big]\,\Omega^{mi}\,\Omega^{nj} \,, \nn\\
  \delta A_\mu{}^{M}    &=& 2 \, \Big[i\,\Omega^{ik}\,\bar\epsilon_k
  \psi_\mu{}^j + \bar\epsilon_k \gamma_\mu \chi^{ijk}\Big] \,  
      {\cal V}_{ij}{}^{M}  \,, \nn\\ 
  \delta B_{\mu\nu\,M} &=& \frac{4}{\sqrt5}\, {\cal V}_M{}^{ij}
       \Big[2\,\bar\psi_{[\mu \,i}\gamma_{\nu]} \epsilon^k\,\Omega_{jk} 
      -i\, \bar\chi_{ijk}  \gamma_{\mu\nu}\epsilon^k \Big]  +2\, d_{MNP}\,
  A_{[\mu}{}^N\,\delta A_{\nu]}{}^P \,, \nn\\
  \delta \psi_\mu{}^i &=& 
     (\pa_\mu \delta^i{}_j - {\cal Q}_{\mu\,j}{}^i - \ft14 \omega_\mu{}^{ab}
  \,\gamma_{ab}\,\delta^i{}_j )\epsilon^j \nn\\
  &&{}   + i\, \Big[\ft1{12}\Big( \gamma_{\mu\nu\rho}\,{\cal H}^{\nu\rho\,ij} 
  -4\, \gamma^\nu\,{\cal H}_{\mu\nu}{}^{ij} \Big)   - g\,\gamma_\mu\,
  A_1{}^{ij} \Big]  \Omega_{jk} \,\epsilon^k  \,, \nn\\
  \delta \chi^{ijk}  &=&{} \ft12 i\, \gamma^\mu\,{\cal P}_\mu{}^{ijkl}
  \,\Omega_{lm} 
  \,\epsilon^m  
    - \ft3{16}\gamma^{\mu\nu}\,\Big[{\cal H}_{\mu\nu}{}^{[ij} \,\epsilon^{k]}
  -\ft13 \Omega^{[ij} \,{\cal H}_{\mu\nu}{}^{k]m} \,\Omega_{mn}\,\e^n \Big]
  \nn\\ 
  &&{} + g\, A_2{}^{l,ijk}\, \Omega_{lm}\,\epsilon^m \,,  
\end{eqnarray}
where 
\begin{equation}
  \label{eq:usp-H}
 {\cal H}_{\mu\nu}{}^{ij} = {\cal H}_{\mu\nu}{}^M\,\vv_M{}^{ij}= ({\cal
 F}_{\mu\nu}{}^M + g\, Z^{MN}  B_{\mu\nu\,N})\, \vv_M{}^{ij} \,.
\end{equation}
Note that $\delta B_{\mu\nu\,M}$ is only determined up to terms that vanish
upon contraction with $Z^{MN}$. While the covariance with respect to most
bosonic 
symmetries is straightforward, the form of $\delta B_{\mu\nu\,M}$ requires
further comment. On $A_\mu{}^M$, the commutator of a supersymmetry
transformation and a vector gauge transformation does not vanish but leads
to, 
\begin{equation}
  \label{eq:vector-susy}
  {[\delta(\epsilon),\delta_{\rm vector}(\Lambda^M)]}= \delta_{\rm tensor}
  (\Xi_{\mu\,M})  \,, 
\end{equation}
with 
\begin{equation}
  \label{eq:Xi=epsilonLambda}
  \Xi_{\mu \,M} = - d_{MNP}\, \Lambda^N\,\delta(\epsilon)A_\mu{}^P\,,
\end{equation}
whereas supersymmetry commutes with tensor gauge transformations. The latter 
requires the presence of the second term in $\delta B_{\mu\nu\,M}$,
proportional to the 
invariant tensor $d_{MNP}$. After this one verifies that the commutator 
(\ref{eq:vector-susy}) is also  correctly realized on the tensor fields (up to
terms that vanish upon contraction with $Z^{MN}$), so
that all supersymmetry variations are consistent
with the bosonic symmetries. What remains is to determine the various
multiplicative coefficients from the requirement that the supersymmetry
algebra closes on all fields. 
With the coefficients adopted in (\ref{eq:susy-algebra}), the commutator of
two supersymmetry transformations with parameters $\epsilon_1$ and
$\epsilon_2$ closes uniformly into the bosonic symmetries,
\begin{equation}
  \label{eq:susy-algebra}
  {[\delta(\epsilon_1),\delta(\epsilon_2)]} = \xi^\mu\,D_\mu + \delta_{\rm
  Lorentz}(\epsilon^{ab})  + \delta_{{\rm USp}(8)}(\Lambda_i{}^j) +
  \delta_{\rm vector}(\Lambda^{M})  + \delta_{\rm  tensor}(\Xi_{\mu\,M}) \,.
\end{equation}
Here $\xi^\mu\,D_\mu$ denotes a covariant general coordinate transformation
with parameter $\xi^\mu$. Such a transformation consists of a spacetime
diffeomorphism with parameter $\xi^\mu$ combined with gauge transformations
with parameters,
\begin{eqnarray}
  \label{eq:cov-gct}
  \epsilon^{ab} &=& -\xi^\mu \,\omega_\mu{}^{ab} \,,\nn\\
  \Lambda_i{}^j &=& - \xi^\mu \,{\cal Q}_{\mu\,i}{}^j \,,\nn\\
  \Lambda^{M} &=& -\xi^\mu \,A_\mu{}^{M} \,,\nn\\
  \Xi_{\mu\,M} &=& - \xi^\nu \,B_{\nu\mu\,M} \,, 
\end{eqnarray}
where
\begin{equation}
  \label{eq:xi}
  \xi^\mu = \ft12 \bar\epsilon_{2i}\gamma^\mu\epsilon_{1}{}^i\,.
\end{equation}
Beyond the covariant general coordinate transformations there are the
symmetry variations indicated explicitly in (\ref{eq:susy-algebra}) with
parameters (up to higher-order fermion terms),
\begin{eqnarray}
  \label{eq:parameters}
  \epsilon_{ab} &=& \ft1{12} i\bar\epsilon_{2i} \gamma_{abcd} \epsilon_1{}^k\,
  \Omega_{jk} {\cal H}^{cdij}  +\ft1{3} i\bar\epsilon_{2i} \epsilon_1{}^k
  \,\Omega_{jk} \,{\cal H}_{ab}{}^{ij}  
  - i g\,\bar\epsilon_{1i}\gamma_{ab}\epsilon_2{}^j\,\Omega_{jk}\,
  A_1{}^{ik}\,, \nn\\[1ex] 
  \Lambda_i{}^j  &=& -  g\, T_i{}^{jkl} \, z_{kl}\,,  \nn\\[1ex]  
  \Lambda^{M} &=&   z^{ij}\, 
  {\cal V}{}_{ij}{}^{M}\,,  \nn\\
  \Xi_{\mu\,M}     &=&  - \frac{8}{\sqrt5}\,\xi_{\mu\,ij}\,
  {\cal V}^{ij}{}_{M} - d_{MNP}\,A_\mu{}^N\, z^{ij}\,{\cal
  V}_{ij}{}^P   \,, 
\end{eqnarray}
where we also used the bilinears $z^{ij}$ and $\xi_\mu{}^{ij}$ which are
pseudoreal and antisymmetric in $[ij]$ and under the interchange of the spinor 
parameters $\epsilon_1$ and $\epsilon_2$,
\begin{eqnarray}
  \label{eq:2}
  z^{ij} &=& -2i\,\Omega^{k[i}\,\bar\epsilon_{2k} \,\epsilon_1{}^{j]}\,, \nn\\
  \xi_{\mu\,ij} &=& \ft12
   \bar\epsilon_{2[i}\gamma_\mu \epsilon_1{}^{k}\,\Omega_{j]k}
  \,,  
\end{eqnarray}
Observe that $\Omega^{ij}\,\xi^\mu{}_{ij}=-\xi^\mu$.  

The closure holds modulo the field equations for the tensor field (which are
linear in derivatives), 
\begin{equation}
  \label{eq:B-field-eq}
  3\, D_{[\mu}{\cal H}_{\nu\rho]}{}^M  -  i
   \ft{1}{\sqrt{5}}\,e\,\varepsilon_{\mu\nu\rho\sigma\tau} \,Z^{MN}\, 
  \vv_N{}^{ij} \,\vv_P{}^{kl}
   \,\Omega_{ik}\,\Omega_{jl} {\cal H}^{\sigma\tau\,P} =0\,.
\end{equation}
Since we are only presenting the results up to higher-order fermion terms, the
field equations for the fermions do not enter at this stage.

In the above we made heavy use of the results derived in
sections~\ref{vector-tensor} and \ref{T-tensor}.   
It is now somewhat tedious but straightforward to derive
the full Lagrangian. We present it up to terms quartic in the
fermion fields (the latter are expected to be independent of the gauge
coupling constant),
\begin{eqnarray}
  \label{eq:Lagrangian}
  e^{-1}{\cal L}&=& -\ft12 R - \ft12 \bar\psi_{\mu i}
  \g^{\mu\nu\rho}\,D_\nu\psi_\rho{}^i - \ft1{16}
  {\cal H}_{\mu\nu}{}^{ij} \,{\cal H}^{\mu\nu}{}^{kl}
  \,\Omega_{ik}\,\Omega_{jl} -\ft23 \bar \chi_{ijk} \,D\!\slash 
 \chi^{ijk}  \nn\\[1ex]
  &&{}
  -\ft1{12} \vert{\cal P}_\mu{}^{ijkl}\vert^2 + \ft23 i\, 
  {\cal P}_\mu{}^{ijkl} \,
  \bar\chi_{ijk} \g^\nu\gamma^\mu \psi_\nu{}^m \,\Omega_{lm} \nn\\[1ex]
  &&{}
  +{\cal H}^{\rho\sigma\,ij} \Big[\ft1{8} i\, \bar\psi_{\mu
  \,i}\gamma^{[\mu} \gamma_{\rho\sigma}\gamma^{\nu]} \psi_\nu{}^k\,
  \Omega_{kj} - \ft1{4}  
  \bar\chi_{ijk} \gamma^\mu \,\gamma_{\rho\sigma} \psi_\mu{}^k   
   -\ft1{2} i\, \bar \chi_{ikl}\g_{\rho\sigma} \chi^{mkl}\,\Omega_{mj} \Big]
  \nn\\[1ex] 
&&{}
+ \frac{\sqrt5}{64\,e} i \varepsilon^{\m\n\rho\sigma\tau}\Big\{ 
 gZ^{MN}
 B_{\m\n\,M} \Big[ D_\rho B_{\sigma\tau\,N} +4\, d_{NPQ} \,A_\rho{}^P
 \Big(\partial_\sigma A_{\tau}{}^Q +\ft13 g\, X_{[RS]}{}^Q\, A_\sigma{}^R
 A_{\tau}{}^S\Big)\Big] \nn\\ 
&&{}{\hspace{11mm}} - \ft83d_{MNP}\Big[ A_\mu{}^M\,\partial_\nu 
   A_\rho{}^N \,\partial_\sigma A_\tau{}^P \nn\\
&&{}{\hspace{23mm}}+ \ft34 g\, X_{[QR]}{}^M \, A_\m{}^N A_\n{}^Q A_{\rho}{}^R
 \Big(\partial_\sigma  A_{\tau}{}^P+\ft15g\, 
 X_{[ST]}{}^P A_\sigma{}^SA_\tau{}^T \Big)\Big] \Big\}\nonumber \\[1ex]
&&{} 
  - \ft32 i  g\, A_1{}^{ik}\Omega_{kj}\,  \bar \psi_{\mu\,i} \g^{\mu\nu}
  \psi_\nu{}^j   
   - \ft43  g \,\Omega_{ml}\, A_{2}{}^{l,ijk}\,\bar{\chi}_{ijk} 
  \gamma^{\mu}\psi_{\mu}{}^{m}  \nn\\[1ex]
  &&{}
  +2 g\,i \, \Omega_{kp}\,\Omega_{lq} \Big[- 4\,A_2{}^{i,jpq}
    + A_1{}^{[i[p}\,\Omega^{q]j]}\Big] \,\bar{\chi}_{ijm}\,\chi^{klm}  
   \nn\\[1ex]
  &&{}
  + \, g^2 \Big[ 3 \,\vert A_{1}{}^{ij}\vert^2 
  - \ft13\,\vert A_{2}{}^{i,jkl}\vert^2 \Big]
 \,,
\end{eqnarray}
where apart from the supersymmetry transformations we made use of the
properties derived for the $T$-tensor derived in section~\ref{T-tensor}. 
The covariant derivatives on the spinor fields are defined by 
\begin{eqnarray}
  \label{eq:D-psi-chi}
  D_\mu\psi_\nu{}^i &=& \pa_\mu \psi_\nu{}^i - 
  {\cal Q}_{\mu\,j}{}^i \,\psi_\nu{}^j 
  - \ft14 \omega_\mu{}^{ab}\,\gamma_{ab}\,\psi_\nu{}^i   \,,\nn\\
  D_\mu\chi^{ijk}  &=&  \pa_\mu \chi^{ijk} -3\, {\cal Q}_{\mu\,l}{}^{[i}
  \,\chi^{jk]l}  - \ft14 \omega_\mu{}^{ab}
  \,\gamma_{ab}\,\chi^{ijk}  \,.
\end{eqnarray}

For the convenience of the reader we record the supersymmetry variations,  
\begin{eqnarray}
  \label{eq:delta-P}
  \delta{\cal P}_{\mu}{}^{ijkl} &=& i D_\mu( 4\, \bar\epsilon_m 
  \chi^{[ijk}\,\Omega^{l]m}  + 3\, \bar\epsilon_m \chi^{m[ij} \,\Omega^{kl]})    
  \nonumber \\ 
  &&{}
  - 2g\, T^{ijkl}{}_{mn} ( i\,\Omega^{mp}\,\bar\epsilon_p
  \psi_\mu{}^n + \bar\epsilon_p \gamma_\mu \chi^{mnp}) \,,\nn \\[1.5ex]
  \delta{\cal H}_{\mu\nu}{}^M&=& 4 D_{[\mu}
  \Big[(i\,\Omega^{ik}\,\bar\epsilon_k  
  \psi_{\nu]}{}^j + \bar\epsilon_k \gamma_{\nu]} \chi^{ijk}) \,  
      {\cal V}_{ij}{}^{M} \Big]  \nn\\ 
   &&{}
    + \frac{4\,g}{\sqrt5} Z^{MN} {\cal V}_N{}^{ij}
       \big(2\,\bar\psi_{[\mu \,i}\gamma_{\nu]} \epsilon^k\,\Omega_{jk} 
      -i\, \bar\chi_{ijk}  \gamma_{\mu\nu}\epsilon^k \big)  \,,
\end{eqnarray}
which are needed for establishing the invariance of the action. The change  of
the scalar potential under (\ref{eq:V-variation}) is also needed,
\begin{eqnarray}
  \label{eq:delta-A-square}
&&{} 
 \delta \Big[ 3 \,\vert A_{1}{}^{ij}\vert^2 
  - \ft13\,\vert A_{2}{}^{i,jkl}\vert^2 \Big]  \nonumber\\
&&{}
\qquad\qquad
= \Big(\ft43 A_1{}^{mq} \,A_{2\,m,ijk}\,\Omega_{lq} 
 +2\,  A_{2}{}^{m,npq} \, A_{2\,n,mij}\,\Omega_{pk}\,
 \Omega_{lq}\Big)\Sigma^{ijkl}   \,,
\end{eqnarray}
which also reveals the condition for stationary points of the 
potential. Here we made use of the fact that the potential is ${\rm USp}(8)$
invariant, so that we can expand in terms of ${\rm USp}(8)$-covariant
variations of the scalar fields, using (\ref{eq:P-Q}). In this way we can also
express the square of the masses at the stationary point, which are then also
proportional to $g^2$ times the square of the $T$-tensor. This pattern repeats
itself for the other fields and all mass squares are simply determined by
expressions quadratic in the $T$-tensor taken at the stationary point. For the
fermions this is already obvious, as the masslike terms are proportional to
the $T$-tensor. For the vector and tensor fields there is a subtlety, as there
are  mixing terms between these types of fields. The mass terms for these
fields gives rise for the following expressions
\bea
({\cal M}^2_{\rm vector})_{MN} &\propto& g^2 \,
{\cal V}_{M}{}^{ij} \,{\cal V}_{N}{}^{kl} \,T^{mnpq}{}_{ij} \,
T^{rstu}{}_{kl}\,\Omega_{mr}\Omega_{ns}\Omega_{pt}\Omega_{qu} \;, 
\nonumber\\[1ex]
({\cal M}_{\rm tensor})^{MN} &\propto& g^2\,{\cal V}_{ij}{}^{M}\,{\cal 
V}_{kl}{}^{N}\,
{\cal Z}^{ij,mn}\,{\cal Z}^{kl,pq}\,\Omega_{mp}\Omega_{nq}
\;.
\eea
Note that the mass term for the tensor fields should not be interpreted as the
mass square, in view of the fact that the kinetic terms are linear in
spacetime derivatives and proportional to $g\,Z^{MN}$. But the result shows
that (when properly taking into account the corresponding kinetic terms) the
vector masses are encoded as eigenvalues of the matrix
$T^{mnpq}{}_{ij} \,
T^{rstu}{}_{kl}\,\Omega_{mr}\Omega_{ns}\Omega_{pt}\Omega_{qu}$, while the
tensor masses correspond to the eigenvalues of ${\cal Z}^{mn,pq}\,$. These
mass terms are 
subject to an orthogonality relation in view of equation~(\ref{eq:z-T}), which
is crucial for obtaining the correct degrees of freedom. To deal with the
mixing between vector and tensor fields, it is best to impose a suitable
gauge. This will be briefly discussed in subsection~\ref{gaugefixing}. 

Stationary points of the potential may lead to a (partial) breaking of
supersymmetry. The residual supersymmetry of the corresponding solution
(assuming maximally symmetric spacetimes) is parametrized by
spinors~$\epsilon^i$ satisfying the condition
\begin{equation}
A_2{}^{l,ijk}\, \Omega_{lm}\,\epsilon^m = 0 \;.
\label{susy_groundstate}
\end{equation}
{}From the gravitino variation one derives an extra condition
\bea
A_1{}^{im}A_{1\,jm} \,\epsilon^j
  = 
   \ft18 (\vert A_1{}^{kl}\vert^2 -\ft19 \vert A_2{}^{k,lmn}\vert^2
   )\,\epsilon^i 
   \;,
   \label{susy_groundstate2}
\eea
but the two conditions~(\ref{susy_groundstate}) 
and~(\ref{susy_groundstate2})
are in fact equivalent by virtue of the second equation 
of~(\ref{eq:A-square}).

\section{Examples}
\setcounter{equation}{0}
In this section we demonstrate our method to the known and some new examples
of maximal ${D=5}$ supergravities. 
These include the original ${\rm SO}(p,q)$ gaugings
of~\cite{GunaRomansWarner}, the ${\rm CSO}(p,q,r)$ gaugings discussed
in~\cite{AndCordFreGual} and the Scherk-Schwarz gaugings
of~\cite{dWST1} (see also \cite{Hull:2003kr,Andrianopoli:2004xu}).
In these examples the gauge group is contained either in the  maximal
${\rm SL}(2,\mathbb{R})\times {\rm SL}(6,\mathbb{R})$ subgroup
of \Exc6, or in a non-semisimple extension of ${\rm SO}(5,5)\times {\rm
  SO}(1,1)$, which is another maximal subgroup of ${\rm E}_{6(6)}$.
Our construction provides a Lagrangian formulation of all these gaugings.
Before coming to the examples we will first discuss the possible gauge fixing
of the tensor gauge transformations in order to make contact with
previous results in the literature. For zero gauge coupling all the tensor
fields disappear from the Lagrangian, and one recovers the result of
\cite{cremmer80} without further ado. 

\subsection{Gauge fixing}
\label{gaugefixing}
The five-dimensional gauged supergravities that so far have appeared in the
literature, have all been formulated without the freedom of tensor gauge
transformations. 
They are recovered from our general formulation by using the tensor gauge
transformations to set some of the vector fields to zero.
To describe this we employ the special basis introduced in
subsection~\ref{specialbasis} and decompose the vector indices according to
$V_M=(V_A,\,V_a,\,V_u)$. By definition, the matrix $Z^{{MN}}$ is invertible on
the space spanned by $V_u$, so that on this space we may define its inverse
$Z_{{uv}}$ according to
\bea
Z_{{uw}}\,Z^{{vw}} &\equiv& \delta_{u}^{v}  \;.
\eea
By means of the tensor gauge transformations in (\ref{eq:A-var}) we  
impose the gauge condition $A_{\mu}{}^{u}= 0$, thus breaking ${\rm E}_{6(6)}$
covariance. Therefore we have to add a compensating term to the supersymmetry
variations,
\begin{equation}
  \label{eq:compensating}
  \delta^{\rm new}(\epsilon) = \delta^{\rm old}(\epsilon) +
  \delta(\Xi_{\mu\,u}) \,,
\end{equation}
with $\Xi_{\mu\,u} = - 2g^{-1} Z_{uv}\,{\cal
  V}^v{}_{ij}\,(i\,\Omega^{ik}\,\bar\epsilon_k \psi_\mu{}^j + \bar\epsilon_k
  \gamma_\mu \chi^{ijk})$. 
The terms proportional to the inverse gauge coupling constant can be
avoided by making a field redefinition. New tensor fields are defined by 
${\cal B}_{\mu\nu}{}^u \equiv {\cal H}_{\mu\nu}{}^u$, which are invariant
under the tensor transformations, just as the field strengths   
${\cal F}_{\mu\nu}{}^A ={\cal H}_{\mu\nu}{}^A$ and 
${\cal F}_{\mu\nu}{}^a ={\cal H}_{\mu\nu}{}^a$. Hence the new tensor fields are
\begin{equation}
  \label{eq:new-B}
{\cal B}_{\mu\nu}{}^{u}  = 
gZ^{uv}\, B_{\mu\nu\,v} +C_{[AB]}{}^{u}\,A_{{\mu}}{}^{A}\,A_{\nu}{}^{B}+
C_{Ab}{}^{u}\,A_{{[\mu}}{}^{A}\,A_{\nu]}{}^{b} \;,
\end{equation}
which will now appear in the Lagrangian as massive fields. Under gauge and
supersymmetry transformations they transform according to 
\bea
\delta(\Lambda)\,{\cal B}_{\mu\nu}{}^{u} &=&
-\,\Lambda^{A} (D_{{Av}}{}^{u}\,{\cal B}_{\mu\nu}{}^{v}
-C_{{AB}}{}^{u}\,{\cal F}_{\mu\nu}{}^{B}
-C_{{Ab}}{}^{u}\,{\cal F}_{\mu\nu}{}^{b}) \;,
\nonumber\\[1ex]
\delta(\epsilon)\,{\cal B}_{\mu\nu}{}^{u} &=&
4\, D_{[\mu} \Big[( i\,\Omega^{ik}\,\bar\epsilon_k
  \psi_{\nu]}{}^j + \bar\epsilon_k \gamma_{\nu]} \chi^{ijk}) \,
      {\cal V}_{ij}{}^{u} \Big] \nonumber\\
&&{}
      {}+ \frac{4 \,g}{\sqrt5} \,Z^{uv}\, {\cal V}_v{}^{ij}
       (2\,\bar\psi_{[\mu \,i}\gamma_{\nu]} \epsilon^k\,\Omega_{jk}
      -i\, \bar\chi_{ijk}  \gamma_{\mu\nu}\epsilon^k )\;,
\eea
where the last expression is a special case of the second equation
(\ref{eq:delta-P}). In the Lagrangian these tensor fields appear in the
kinetic term of the modified field strength tensor, 
${\cal H}_{\mu\nu}{}^{ij}={\cal V}_{A}{}^{ij}{\cal F}_{\mu\nu}{}^{A}+
{\cal V}_{a}{}^{ij}{\cal F}_{\mu\nu}{}^{a}+{\cal V}_{u}{}^{ij}{\cal
  B}_{\mu\nu}{}^{u}$, 
and in the Chern-Simons term whose leading term now takes the form
\bea
{\cal L}_{{\rm VT}} &\propto&
\ft12 i \varepsilon^{\m\n\rho\sigma\tau}\, g^{{-1}}\,Z_{uv}\,
 {\cal B}_{\m\n}{}^{u}  D_\rho {\cal B}_{\sigma\tau}{}^{v} + \cdots \;.
\eea
The appearance of the inverse coupling constant $g^{-1}$ and the matrix
$Z_{uv}$ in this term shows that, after gauge fixing, the theory no longer
possesses a smooth limit to the ungauged theory. This phenomenon has been
observed in the original construction of the ${\rm SO}(p,q)$ gauged
theories~\cite{GunaRomansWarner}. 
Note that the full Lagrangian~(\ref{eq:Lagrangian}) in contrast allows
a smooth limit $g\rightarrow0$.

In the gauge-fixed version there remain many more interaction terms between
tensor and vector fields than those that are known from the ${\rm SO}(q,6-q)$
gaugings. These terms have a similar structure as the terms that were found
recently for non-maximal gauged supergravities with eight supersymmetries
\cite{BergshVandoren}. 

\mathversion{bold}
\subsection{${\rm CSO}(p,q,r)$ gaugings}
\mathversion{normal}
Let us first review the case of gauge groups contained in the 
${\rm SL}(2,\mathbb{R})\times {\rm SL}(6,\mathbb{R})$ maximal subgroup
of \Exc6. Recall that a consistent gauging is completely encoded in an
embedding 
tensor~$\Theta_{M}{}^{\alpha}$ that satisfies the linear projection
constraint~(\ref{eq:repres-constraint}) and any of the equivalent forms
of the quadratic constraint~(\ref{eq:equiv-quadr}).
With respect to ${\rm SL}(2,\mathbb{R})\times {\rm SL}(6,\mathbb{R})$,
the representations of the vector gauge fields, the \Exc6
generators and the embedding tensor decompose according to,
\begin{eqnarray}
\label{eq:decblock-sl6}
\overline{\bf 27} &\rightarrow& ({\bf 1},\overline{{\bf 15}}) +
({\bf 2},{\bf 6}) \,,
\nonumber\\
{\bf 78} &\rightarrow&
({\bf 1},{\bf 35}) + ({\bf 3},{\bf 1})
+ ({\bf 2},{\bf 20})\,,
\nonumber\\
{\bf 351} &\rightarrow&
({\bf 1},{\bf 21}) + ({\bf 3},{\bf 15}) + ({\bf 2},\overline{\bf 84})
+ ({\bf 2},\overline{\bf 6}) + ({\bf 1},{\bf 105})\,,
\end{eqnarray}
respectively. A generic embedding tensor $\Theta_{M}{}^{\alpha}$
transforming in the ${\bf 351}$
representation of \Exc6 thus couples vector fields to generators according to
\begin{equation}
\begin{tabular}{|c| c c| } \hline
~&~&~\\[-6mm]
&$({\bf 1},{\bf 15})$
&$({\bf 2},\overline{\bf 6})$
\\ \hline
~&~&~\\[-4.5mm]
$({\bf 1},{\bf 35})$
&$({\bf 1},{\bf 21}) + ({\bf 1},{\bf 105}) $
&$({\bf 2},\overline{\bf 6})+({\bf 2},\overline{\bf 84}) $
\\
$({\bf 3},{\bf 1})$
&$({\bf 3},{\bf 15})$
&$({\bf 2},\overline{\bf 6})$
\\
$({\bf 2},{\bf 20})$
&$({\bf 2},\overline{\bf 6})+({\bf 2},\overline{\bf 84}) $
&$({\bf 3},{\bf 15}) + ({\bf 1},{\bf 105}) $
\\ \hline
\end{tabular}
\end{equation}
Equivalent representations in the bulk of the table must
be identified since all representations in the decomposition of
the ${\bf 351}$ representation appear with multiplicity one.
We stress that the representations indicated in the first row refer to 
the charges, which transform in the ${\bf 27}$ 
representation, and not to the gauge fields which transform in the
$\overline{\bf 27}$ representation. Also the first column refers to the
conjugate representation of the representations into which the \Exc6 generators
decompose, because $\Theta$ carries upper indices $\alpha$ unlike the \Exc6
generators. However, the representations in the first column happen to be
self-conjugate. This will not be the case in our next example.

Searching for subgroups of ${\rm SL}(2,\mathbb{R})\times {\rm
SL}(6,\mathbb{R})$ implies that the representations in the last row must be
excluded. Hence the only possible representation assignment
for the embedding tensor is the representation~$({\bf 1},{\bf 21})$. This
representation
can be described by a symmetric six-by-six tensor $\theta_{IJ}$, where the
indices $I,J= 1,\ldots,6$ denote the vector indices of
${\rm SL}(6,\mathbb{R})$. This
restricts the possible gauge groups to subgroups of ${\rm SL}(6,\mathbb{R})$
and the participating vector gauge fields to the $({\bf 1},\overline{\bf
15})$ representation.  Denoting vector indices of ${\rm
SL}(2,\mathbb{R})$ by $\underline\alpha=1,2$,  the vector fields now decompose
into $A_{\mu}{}^{M}\rightarrow (A_{\mu}{}^{IJ},A_{\mu}{}_{I\underline\alpha})$.
The embedding tensor is then  parametrized in terms
of $\theta_{IJ}$ according to
$\Theta_{[IJ]}{}^{K}{}_{L} = \d^K{}_{[I}\,\theta^{~}_{J]L}$ and all other
components vanish. This leads to the gauge group
generators,
\begin{equation}
  \label{eq:X-CSO}
X_{IJ} = \theta_{L[I}\, t_{J]}{}^{L}\;,
\end{equation}
with $t_{K}{}^{L}$ the ${\rm SL}(6,\mathbb{R})$-generators.
Similarly, one finds that the only nonvanishing components of the
antisymmetric tensor $Z^{MN}$ are given by
\bea
Z_{I{\underline{\alpha}}\,J{\underline{\beta}}} &\propto&
\varepsilon_{{\underline\alpha\underline\beta}}\,\theta_{IJ}\;, 
\label{ZSL6}
\eea
The explicit form of $\Theta_{M}{}^{\alpha}$ and $Z^{MN}$ shows that
$Z^{MN}\Theta_{N}{}^{\alpha}=0$ for any choice of $\theta_{IJ}$.
According to~(\ref{eq:equiv-quadr}), the quadratic constraint is
thus satisfied and every symmetric six-by-six tensor $\theta_{IJ}$
defines a viable gauging. The gauge group is contained in the subgroup of 
${\rm SL}(6,\mathbb{R})$ that leaves $\theta_{IJ}$ invariant.
This can be verified explicitly by making use of (\ref{eq:X-CSO}). 

The ${\bf 21}$ representation associated with $\theta_{IJ}$ falls in
28 different conjugacy classes leading to 15 independent gaugings. The
corresponding tensors take the form,
\bea
\theta_{IJ}&=& {\rm diag}(\,\underbrace{1, \dots,}_{p}\underbrace{-1,
\dots,}_{q} \underbrace{0, \dots}_{r}\,)  \;,
\label{thetaSL6}
\eea
with $p+q+r=6$. The corresponding gauge group is ${\rm CSO}(p,q,r)$. We
have thus obtained a complete classification of possible gauge groups
${\rm G}_g\subset {\rm SL}(2,\mathbb{R})\times {\rm
SL}(6,\mathbb{R})$. From the rank of the tensors $\Theta_M{}^\alpha$
and $Z^{MN}$ one determines the number of tensor fields and the number of
vector fields (after an appropriate gauge choice). It follows that the number
of tensor fields is equal to $t=2(6-r)$, and the number of vector fields that
gauge the group ${\rm CSO}(p,q,r)$ equals $s=\ft12(6-r)(5+r)$. The latter
decompose into the gauge fields associated with the subgroup ${\rm SO}(p,q)$
and with $r(p+q)$ nilpotent generators. Furthermore the
number of abelian gauge fields equals $\ft12 r(r-1)$. A Lagrangian formulation
for these non-semisimple groups had not yet been obtained. It now
follows directly from the universal Lagrangian~(\ref{eq:Lagrangian}).

\subsection{Gaugings characterized by ${\rm SO}(5,5)\times {\rm SO}(1,1)
  \subset {\rm E}_{6(6)}$ }
Another class of gaugings is based on the decomposition of ${\rm
E}_{6(6)}$ under its subgroup ${\rm SO}(5,5)\times {\rm SO}(1,1)$, where the
first factor is the U-duality group of maximal supergravity in
six dimensions. The decompositions of the representations of the vector gauge
fields, the \Exc6 generators and the embedding tensor are now given by,
\begin{eqnarray}
\overline{\bf 27} &\to& \overline{\bf 16}_{-1}+{\bf 10}_{+2}+{\bf
1}_{-4}\,,\nonumber\\
{\bf 78} &\to& {\bf 45}_{0}+{\bf 1}_{0}+ {\bf 16}_{-3}+\overline{{\bf
16}}_{+3}\,,\nonumber\\
{\bf 351} &\to& {\bf 144}_{+1}+
{\bf 16}_{+1}+{\bf 45}_{+4}+{\bf 120}_{-2}+{\bf 10}_{-2}+{\bf
\overline{16}}_{-5}\,.
\end{eqnarray}
Hence we effect the decomposition of the vector fields by assigning
vector indices $m,n=1, \dots, 10$ and spinor indices $\underline \alpha=1,
\dots, 16$, with respect to ${\rm SO}(5,5)$, respectively. The gauge
fields then decompose according to 
$A_{\mu}{}^{M}\rightarrow (A_{\mu}{}^{\underline\alpha},
A_{\mu}{}^{m},A_{\mu}{}^{0})$ 
whereas the ${\rm E}_{6(6)}$ generators decompose according to $t_\alpha \to (t_{mn}, t_0,
  t_{\underline\alpha}, t^{\underline\alpha})$. 

Upon extending ${\rm SO}(5,5)\times {\rm SO}(1,1)$ with the 16 nilpotent
generators belonging to either the ${\bf 16}_{-3}$ or the $\overline{\bf
  16}_{+3}$ representation, the resulting non-semisimple group constitutes a
maximal subgroup of \Exc6. The gauge couplings induced by a generic embedding
tensor $\Theta_{M}{}^{\alpha}$ transforming in the ${\bf 351}$ representation
of \Exc6 are as follows, 
\begin{equation}
\label{tableSO}
\begin{tabular}{|c|ccc|} \hline
&${\bf 10}_{-2}$
&${\bf 16}_{+1}$
&${\bf 1}_{+4}$
\\
\hline
${\bf 16}_{-3}$
&${\overline{\bf 16}}_{-5} $
&${\bf 120}_{-2}+ {\bf 10}_{-2}$
&${\bf 16}_{+1}$
\\
${\bf 45}_{0}$
&${\bf 10}_{-2}+{\bf 120}_{-2}$
& ${\bf 144}_{+1}+ {\bf 16}_{+1}$
& ${\bf 45}_{+4}$
\\
${\bf 1}_{0}$
&${\bf 10}_{-2}$
&${\bf 16}_{+1}$
&
\\
$\overline{{\bf16}}_{+3}$
&${\bf 144}_{+1}+ {\bf 16}_{+1}$
&${\bf 45}_{+4}$
&\\
\hline
\end{tabular}
\end{equation}
Again equivalent representations for the embedding matrix are
identified as they appear with multiplicity one in the decomposition of the
${\bf 351}$ representation. Note again that the first row denotes the
representation of the charges and not of the gauge fields, whereas the
assignment in the first column denotes the conjugate representation as
compared to the corresponding \Exc6 generators. In this way, the upper-left
entry thus describes the coupling of the gauge fields in the ${\bf 10}_{+2}$ to
the generators in the $\overline{\bf 16}_{+3}$ generators.

{}From the table one immediately concludes that no
subgroup of ${\rm SO}(5,5)\times {\rm SO}(1,1)$ can be gauged consistently, as
there is no irreducible component of the embedding tensor that appears
exclusively in the two middle rows of the table. We will therefore search for
gauge groups that also involve (nilpotent) generators from either the
${\bf {16}}_{-3}$ or the ${\bf \overline{16}}_{+3}$ representation.

Let us start with those gaugings that couple generators belonging to the  
${\bf {16}}_{-3}$ representation. According to table~(\ref{tableSO}) this
allows two irreducible components for the embedding tensor, namely the 
${\bf 45}_{+4}$ and ${\bf 144}_{+1}$ representations. 
We first focus on the case of an embedding tensor in the ${\bf 45}_{+4}$
representation. The corresponding embedding tensor is parametrized in terms
of an antisymmetric ten-by-ten tensor $\theta^{mn}$ according to
\bea
\Theta_{\underline\alpha}{}^{\underline\beta} \propto 
   \theta^{mn} \,(\Gamma_{mn})_{\underline\alpha}{}^{\underline\beta}
   \;,\qquad
     \Theta_0{}^{mn} \propto \theta^{mn}\;,
\eea
where the ${\rm SO}(5,5)$ generators in the (chiral) spinor representation are
denoted by $(\Gamma_{mn})_{\underline\alpha}{}^{\underline\beta}$. 
The only nonvanishing components of the tensor $Z^{MN}$ are given by
\bea
Z^{mn} \propto \theta^{mn} \;.
\eea
Together, this implies that only (some of the) vector fields
$A_{\mu}{}^{\underline\alpha}$ 
and $A_{\mu}{}^{0}$  from the ${\bf \overline{16}}_{-1}+{\bf 1}_{-4}$
representation can participate in the gauging and, furthermore, that only
(some of the) tensor fields $B_{\mu\nu}{}^{m}$ from the ${\bf 10}_{-2}$
representation will remain in the gauge-fixed formulation. Clearly
$Z^{MN}\Theta_{N}{}^{\alpha}=0$ for any choice of $\theta^{mn}$, so that the  
quadratic constraint~(\ref{eq:equiv-quadr}) is satisfied
and every antisymmetric ten-by-ten tensor $\theta^{mn}$ defines
a viable gauging. The theories descending from $D=6$
dimensions by Scherk-Schwarz reduction belong to this class~\cite{dWST1}, with
the tensor $\theta^{mn}$ 
singling out the generator of  ${\rm SO}(5,5)$ that is associated with
the compactified sixth dimension in the reduction.

On the other hand an embedding tensor $\Theta_M{}^\alpha$ living in the 
${\bf 144}_{+1}$ is parametrized by a tensor $\theta_m{}^{\underline\alpha}$
subject to $(\Gamma^m)_{\underline\alpha\underline\beta}\,
\theta_m{}^{\underline\alpha}=0$, according to 
\begin{eqnarray}
   \label{eq:theta-144}
   \Theta_m{}^{\underline\alpha}\propto  \theta_m{}^{\underline\alpha} \;,
   \qquad 
   \Theta_{\underline\alpha}{}^{mn} \propto
   (\Gamma^{[m})_{\underline\alpha\underline \beta} \,
\theta^{n] \underline \beta}\;,\qquad
   Z^{{m\underline \alpha}} =-Z^{{\underline{\alpha} m}} \propto 
  \theta^{m\underline \alpha}\;.
\end{eqnarray}
Here $(\Gamma^m)_{\underline\alpha\underline\beta}$ is symmetric in the spinor
indices $\underline\alpha, \underline\beta$ and corresponds to the ${\rm
  SO}(5,5)$  gamma matrices restricted to the chiral subspace (after
multiplying with the charge conjugation matrix). 
In this case, the quadratic constraint~(\ref{eq:equiv-quadr}) implies the
nontrivial relations, 
\bea
\theta_m{}^{\underline\alpha} \,\theta^{m\,\underline\beta} =0\;,\qquad
(\Gamma^{[m})_{\underline\alpha\underline\beta} \, 
  \theta^{n]\underline\beta}\,\theta^{k\underline\alpha} =0 \;,
\eea
to be satisfied by $\theta_m{}^{\underline\alpha}$.
It is obvious that solutions to these constraints will correspond (by
dimensional reduction) to maximal gauged supergravities in six dimensions
\cite{dWST1}, because for that theory the embedding tensor belongs to the 
${\bf 144}_{+1}$ representation of the ${\rm SO}(5,5)$ duality group. A
particular  
solution is obtained by restricting $\theta_m{}^{\underline\alpha}$ to the
unique component that is invariant under the diagonal ${\rm SO}(5)$ subgroup
of ${\rm SO}(5)\times{\rm SO}(5)\subset {\rm  SO}(5,5)$. The assignment of
$\theta_m{}^{\underline\alpha}$ with respect to ${\rm SO}(5)$ follows from the
observation that the vector and spinor representations decompose
 according to ${\bf 10}\to {\bf 5}+{\bf 5}$ and ${\bf 16}\to 
{\bf 1} + {\bf 5} + {\bf 10}$, respectively. This leads to one
singlet for $\theta_m{}^{\underline\alpha}$ in view of the fact that 
$\theta_m{}^{\underline\alpha}$ is traceless upon contracting with 
${\rm SO}(5,5)$ gamma matrices. This particular choice for 
the embedding tensor must thus be related by dimensional reduction to the
six-dimensional ${\rm SO}(5)$  gauging~\cite{Cowdall}
(which in turn can be obtained by dimensional reduction from a
seven-dimensional gauging). It corresponds to a
gauging of ${\rm CSO}(5,0,1)$. Indeed, the embedding tensor in the previous
subsection contains precisely one ${\rm SO}(5)$ singlet (under ${\rm SO}(5)$
the ${\bf 21}$ representation of ${\rm SO}(6)$ decomposes into 
${\bf 15}+{\bf 5}+ {\bf 1}$), corresponding to 
$p=5$, $q=0$ and $r=1$. This gauging involves 10 gauge fields associated with
${\rm SO}(5)$ and 5 extra gauge fields from the 
${\bf 10}_{+2}$ representation.

Obviously, there are also gaugings in which the embedding tensor has
nonvanishing components in both the ${\bf 45}_{+4}$ and the ${\bf 144}_{+1}$
representations. According to~(\ref{eq:equiv-quadr}) this implies the
additional identities, 
\bea
\theta^{mn}\,\theta_n{}^{\underline\alpha}=0\;,\qquad
\theta^{mn}\,\theta^{k {\underline\alpha}}\,
(\Gamma_{mn})_{\underline\alpha}{}^{\underline\beta}=0\;, 
\eea
among the different components of $\Theta_{M}{}^{\alpha}$. These gaugings
will include, for example, the theories obtained by a two-fold Scherk-Schwarz
reduction from seven dimensional supergravity.

Finally, let us briefly consider the class of gaugings that include gauge
group generators from the ${\bf \overline{16}}_{+3}$ representation, so that
the gauge group is contained in the conjugate extension of 
${\rm SO}(5,5)\times {\rm SO}(1,1)$ to a maximal subgroup of \Exc6. 
According to table~(\ref{tableSO}), allowed embedding tensors have now
components in the ${\bf 10}_{-2}$, 
the ${\bf \overline{16}}_{-5}$, and the ${\bf 120}_{-2}$ representations. 
A gauging of the first type is parametrized by a constant vector $\theta_{m}$
with nontrivial components, 
\bea
\Theta_{m}{}^{kl}&=& \delta_{m}^{[k}\,\theta^{l]}\;,\qquad
\Theta_{m}{}^{0}~\propto~ \theta_{m}\;,\qquad
\Theta_{\underline\alpha}{}_{\,{\underline\beta}}~\propto~ 
\theta_{m}\,(\Gamma^{m})_{{\underline\alpha}{\underline\beta}}\;,
\nonumber\\
Z^{m0}&=&-Z^{{0m}} ~\propto~ \theta^{m}\;,
\label{g10}
\eea
in $\Theta_{M}{}^{\alpha}$ and $Z^{MN}$.\footnote{Note 
that in $\Theta_{\underline\alpha}{}_{\,{\underline\beta}}$
the index $\underline\alpha$ couples to vector fields in the 
${\bf \overline{16}}_{-1}$ while the index $\underline\beta$ couples to 
generators in the ${\bf \overline{16}}_{+3}$.}
The quadratic
constraint~(\ref{eq:equiv-quadr}) is then equivalent to the condition 
$\theta^{m}\theta_{m}=0$, which does admit nontrivial real solutions. 
Every lightlike vector $\theta_{m}$ thus defines a viable gauging that 
involves only two tensor fields $B_{\mu\nu\,0}$ and
$\theta^{m}B_{\mu\nu\,m}$. 

An embedding tensor in the ${\bf \overline{16}}_{-5}$ representation is
parametrized by a spinor $\theta^{\underline\alpha}$ which induces the
components 
 \bea
 \Theta_{m\,{\underline\alpha}}&=&
 (\Gamma_{m})_{{\underline\alpha}{\underline\beta}}\, 
 \theta^{\underline\beta}\;,   
 \qquad
 Z^{0{\underline\alpha}}~=~-Z^{{{\underline\alpha} 0}}  ~\propto~
 \theta^{\underline\alpha} \;, 
 \label{lag}
 \eea
showing that the quadratic constraint~(\ref{eq:equiv-quadr}) is automatically
satisfied. These gaugings constitute a new class of 
abelian gaugings that involve vector fields
exclusively from the ${\bf 10}_{+2}$ and
only two tensor fields, $B_{\mu\nu\,0}$ and 
$\theta^{\underline\alpha}B_{\mu\nu\,{\underline\alpha}}$. 
In fact, they have a geometrical interpretation originating 
from type-IIB RR-flux compactifications on a five-torus~$T^{5}$.
To work out this relation, representations are further decomposed under
the group ${\rm SL}(5,\mathbb{R})\times {\rm SO}(1,1)$,
associated to the metric moduli of $T^5$
and the Cartan subgroup of the ten-dimensional ${\rm SL}(2,\mathbb{R})$
duality group, respectively.
This ${\rm SO}(1,1)$ is a combination of the two ${\rm SO}(1,1)$
factors appearing in
${\rm SL}(5,\mathbb{R})\times {\rm SO}(1,1)\times {\rm SO}(1,1)\subset
 {\rm SO}(5,5)\times {\rm SO}(1,1)$.
 The embedding tensor $\theta^{\underline\alpha}$
then gives rise to three irreducible components 
${\bf \overline{10}}_{+1}$, ${\bf 5}_{+2}$, ${\bf 1}_{0}$,
corresponding to a three-form, a one-form, and a five-form RR-flux,
respectively
\begin{eqnarray}
\partial^{\vphantom{(l)}}_{[\Lambda}C^{(2)}_{\Sigma\Gamma]}&\propto&
\theta_{\Lambda\Sigma\Gamma}
\;,
\qquad
\partial_{\Lambda}C^{(0)}~\propto~ \theta_\Lambda
\;,
\qquad
\partial^{\vphantom{(l)}}_{[\Lambda}C^{(4)}_{\Sigma\Gamma\Delta\Pi]}
~\propto~\epsilon_{\Lambda\Sigma\Gamma\Delta\Pi}\,\theta
\;,
\end{eqnarray}
where indices $\Lambda,\Sigma,\dots$ refer to coordinates on the five-torus
and $C^{(0)}$, $C^{(2)}$, and $C^{(4)}$ denote the RR-fields in ten dimensions.
After gauge fixing, the vector fields can be assigned the representations
\bea
(B^{(2)}_{\mu \Lambda},G^{\Lambda}_\mu)~=~ {\bf 5}_{-1}+\overline{{\bf 5}}_{0}
~\subset~ {\bf 10}_{+2}
\;,
\quad
(C^{(4)}_{\mu\Lambda\Sigma\Gamma},C^{(2)}_{\mu\Sigma})~=~
\overline{{\bf 10}}_{0}+{\bf 5}_{+1}~\subset~ {\bf\overline{16}_{-1}}
\;.
\eea
{}From (\ref{lag}) it follows, that scalars 
in the presence of these fluxes couple only to graviphotons $G^{\Lambda}_\mu$
and vector fields originating from the NSNS two form $B^{(2)}$.
The two tensor fields in turn descend from $B^{(2)}$ and $C^{(2)}$.
Details can be worked out along the lines of~\cite{dWST3}.

\section{Conclusions}
In this paper we presented deformations of maximally supersymmetric
$D=5$ supergravity induced by gauge interactions. No other 
supersymmetric deformations of this theory are expected to exist. The deformed
theory is described by the 
Lagrangian~(\ref{eq:Lagrangian}) together with the supersymmetry
transformation rules~(\ref{eq:susy-trans}). This Lagrangian gives a uniform 
description of all possible deformations in a manifestly 
${\rm E}_{6(6)}$-covariant framework. 
It couples vector fields in the $\overline{\bf 27}$ and tensor
fields in the ${\bf 27}$ representation of \Exc6, which in an intricate way
transform under vector and tensor gauge transformations according
to~(\ref{eq:A-var}) and (\ref{eq:delta-B}), respectively. As a result the
number of degrees of freedom is always consistent with supersymmetry. 

The gauging is entirely encoded in the constant embedding
tensor $\Theta_{M}{}^{\alpha}$ which belongs to the ${\bf 351}$
representation of ${\rm E}_{6(6)}$ and satisfies the quadratic
constraint~(\ref{eq:equiv-quadr}). It describes the coupling of vector fields
to gauge group generators~(\ref{X-theta-t}) and implies the existence of an
(antisymmetric) metric $Z^{MN}$ that serves as a metric for the first-order
kinetic term of the two-form tensor fields~(\ref{eq:CS-term}), which is
accompanied by Chern-Simons terms. Also the tensor gauge
transformations depend on the tensor $Z^{MN}$. 
In contrast to the ungauged theory~\cite{cremmer80}, the
Lagrangian of the gauged supergravity  combines both the vector fields
and their dual tensor fields, where the embedding tensor projects out those
vector and tensor fields that actually participate in the gauging.
This formulation admits a smooth limit $g\rightarrow0$ back to the 
ungauged theory.
In section~\ref{gaugefixing}, we have discussed the form of the
Lagrangian~(\ref{eq:Lagrangian}) after a specific gauge choice
which fixes the freedom of tensor gauge transformations by
eliminating part of the vector fields and explicitly breaks the
${\rm E}_{6(6)}$ covariance. Previous constructions of
gauged supergravities in five dimensions have been obtained
in this special gauge~\cite{GunaRomansWarner,AndCordFreGual}, with the
exception of the work described in \cite{ortin}, where a variety of
vector-tensor dualities is applied in the presence of Stueckelberg-type
vectors. 

The universal formulation of the five-dimensional gauged supergravity shows a
strong similarity with the formulation of the three-dimensional gauged
supergravities~\cite{NicSam}.
In three dimensions, the relevant duality relates scalar and vector fields and
the ungauged theory is formulated entirely in terms of scalar fields.
The general gauged theory on the other hand combines the complete scalar
sector and the dual vector fields. The latter satisfy
first-order field equations and do not carry additional degrees of freedom.
The gauged Lagrangian is manifestly ${\rm E}_{8(8)}$ covariant and the
embedding tensor is a symmetric matrix in the ${\bf 1}+{\bf 3875}$
representation that projects out the vector fields that actually participate
in the gauging. In close analogy to the five-dimensional case, it describes
the coupling of vector fields to symmetry generators and simultaneously serves
as a metric for the first-order kinetic term of the vector fields, which here
is a standard Chern-Simons term. 

In all spacetime dimensions the embedding tensor is subject to a linear
representation constraint, required by supersymmetry, and a quadratic
constraint to ensure the closure of the gauge algebra. As far as we know,
there are no other conditions to ensure the consistency, irrespective of the
spacetime dimension. In a forthcoming paper \cite{dWST4} we will analyze the
four-dimensional gaugings and present a similar result. In this case  there
are no vector-tensor dualities, but one has to deal with electric/magnetic 
duality. Although the details are quite different and the group-theoretical
analysis proceeds along different lines, the final result is qualitatively the
same and one 
obtains a uniform Lagrangian with the possible gaugings encoded in an
embedding tensor transforming in the ${\bf 912}$ representation of \Exc7. 

We expect this pattern to persist in higher spacetime dimensions as well. For
higher dimensions, one has, however, to cope with a larger variety of tensor
fields. While the duality group \Exc{11-D} becomes more simple, so that the
group theory analysis becomes more straightforward, the structure of the field
representation becomes more complicated. In this respect the seven-dimensional
maximal supergravity theories are an interesting testing ground. Here the
relevant duality relates two- and three-form tensors and 
the ungauged theory is formulated entirely in terms of the two-form
fields~\cite{Sezgin:1982gi}. In analogy to the three-dimensional scenario and
the five-dimensional scenario presented here, one thus expects a universal
Lagrangian for the general seven-dimensional gauged maximal supergravities
that combines the 
two-form fields with their dual three-form tensors. Both these tensors should
be subject to tensor gauge transformations to ensure the correct number of
degrees of freedom. The embedding tensor in seven dimensions contains the
${\bf 15}$ representation of ${\rm E}_{4(4)}={\rm SL}(5)$~\cite{dWST1} and may
act as a (symmetric) metric for a first-order kinetic term of the 3-rank
tensor fields. The latter transform in the $\overline{\bf 5}$ representation. 
This particular embedding tensor leads to all the ${\rm CSO}(p,q,r)$ gaugings
with $p+q+r=5$. Gauge-fixing the rank-3 tensor gauge invariance will reproduce
the known form~\cite{PPvN} of the gauged theory which no longer admits a
smooth limit $g\rightarrow0$ to the ungauged theory. 

However, from the existence of certain Scherk-Schwarz reductions from 
eight-dimen\-sional supergravity, one deduces that the embedding tensor should
in general belong to the ${\bf 15}+{\overline{\bf 40}}$ representation, so
that the assignment originally proposed in \cite{dWST1} will be too
restrictive. This extension of the embedding tensor induces a 
coupling between the vector 
fields and 2-and 3-rank tensor fields, based on a nontrivial extension of the
tensor-vector gauge invariances discussed in this paper. It should
be possible to incorporate these gaugings in the context of a universal
Lagrangian of the type discussed in this paper. We will report on this theory 
and related issues elsewhere \cite{dWST5}. 

Finally, one may wonder what the physical significance could be of the extra
tensor fields that one needs for incorporating certain gaugings in a U-duality
covariant way. From an M-theory perspective the supergravity fields couple to
U-duality representations of BPS states and this coupling may induce the
gauging. Obviously, such couplings could involve certain supergravity fields
which will not necessarily describe dynamical degrees of freedom and which
could be dropped in the limit of vanishing gauge coupling constant. All of
this is reminiscent of the arguments leading to BPS-extended supergravity,
which were presented some time ago \cite{dW-S00,dWNic-G01}. We expect
that the universal Lagrangian constructed here may well have a role
to play in this context.  

\vspace{8mm}
\noindent
{\bf Acknowledgement}\\
\noindent
This work is partly supported by EU contracts HPRN-CT-2000-00131,
MRTN-CT-2004-005104, and MRTN-CT-2004-503369, the INTAS contract 03-51-6346,
and the DFG grant SA 1336/1-1. 

\bigskip

%

%
\end{document}